\documentclass[trackchanges,twocolumn]{aastex701}

\usepackage{amsmath}
\usepackage{booktabs} 
\usepackage{tabularx}   
\usepackage{xspace}
\usepackage{comment}
\usepackage{enumitem}
\usepackage{subcaption}
\usepackage{multirow}

\newcommand{\Msun}{\ensuremath{M_\odot}}
\newcommand{\Rsun}{\ensuremath{R_\odot}}

\newcommand{\Zsun}{\ensuremath{Z_\odot}}
\newcommand{\posydon}{\texttt{POSYDON}\xspace}
\newcommand{\mesa}{\texttt{MESA}\xspace}
\newcommand{\Hecore}{\texttt{He core}\xspace}
\newcommand{\COcore}{\texttt{C/O core}\xspace}
\newcommand{\pixelpop}{\textsc{PixelPop}\xspace}
\newcommand{\fullpop}{\textsc{FullPop}\xspace}

\shorttitle{Cosmological Double Neutron Stars}

\begin{document}

\title{From the Milky Way to the Early Universe: How Metallicity Shapes the Formation of Double Neutron Stars}

\author[0000-0002-6064-388X]{Abhishek Chattaraj}\email{a.chattaraj@ufl.edu}
\affiliation{Department of Physics, University of Florida, 2001 Museum Rd, Gainesville, FL 32611, USA}

\author[0000-0001-5261-3923]{Jeff J. Andrews}\email{jeffrey.andrews@ufl.edu}
\affiliation{Department of Physics, University of Florida, 2001 Museum Rd, Gainesville, FL 32611, USA}
\affiliation{Institute for Fundamental Theory, 2001 Museum Rd, Gainesville, FL 32611, USA}

\author[0000-0003-0648-2402]{Monica Gallegos-Garcia}\email{monica.gallegos_garcia@cfa.harvard.edu}
\affiliation{Harvard Society of Fellows, 78 Mount Auburn Street, Cambridge, MA 02138, USA}
\affiliation{Center for Astrophysics \textbar{} Harvard \& Smithsonian, 60 Garden St. Cambridge, MA, 02138, USA}

\author[0000-0002-6842-3021]{Max M. Briel}\email{max.briel@unige.ch}
\affiliation{Département d’Astronomie, Université de Genève, Chemin Pegasi 51, CH-1290 Versoix, Switzerland}
\affiliation{Gravitational Wave Science Center (GWSC), Université de Genève, CH1211 Geneva, Switzerland}

\author[0000-0002-0147-0835]{Michael Zevin}\email{mzevin@adlerplanetarium.org}
\affiliation{The Adler Planetarium, 1300 South DuSable Lake Shore Drive, Chicago, 60605, IL, USA}
\affiliation{Center for Interdisciplinary Exploration and Research in Astrophysics (CIERA),1800 Sherman, Evanston, IL 60201, USA}
\affiliation{NSF-Simons AI Institute for the Sky (SkAI), 172 E. Chestnut St., Chicago, IL 60611, USA}

\author[0000-0003-1474-1523]{Tassos\,Fragos}\email{Anastasios.Fragkos@unige.ch}
\affiliation{Département d’Astronomie, Université de Genève, Chemin Pegasi 51, CH-1290 Versoix, Switzerland}
\affiliation{Gravitational Wave Science Center (GWSC), Université de Genève, CH1211 Geneva, Switzerland}

\author[0000-0002-7464-498X]{Emmanouil\,Zapartas}\email{ezapartas@ice.csic.es}
\affiliation{Institute of Space Sciences (ICE), CSIC, Campus UAB, Carrer de Can Magrans s/n, E-08193 Barcelona, Spain}

\author[0000-0002-3874-2769]{David R. Aguilera-Dena}\email{daguilera@astro.unam.mx}
\affiliation{Instituto de Astronom\'ia, Universidad Nacional Aut\'onoma de M\'exico, Ap. 70-264, 04510 CDMX, M\'exico}
\affiliation{Department of Astronomy and Astrophysics, University of California, Santa Cruz, CA 95064, USA}

\author[0000-0001-6692-6410]{Seth Gossage}\email{seth.gossage@northwestern.edu}
\affiliation{Center for Interdisciplinary Exploration and Research in Astrophysics (CIERA),1800 Sherman, Evanston, IL 60201, USA}
\affiliation{NSF-Simons AI Institute for the Sky (SkAI), 172 E. Chestnut St., Chicago, IL 60611, USA}

\author[0000-0002-7322-4748]{Anarya Ray}\email{anarya.ray@northwestern.edu}
\affiliation{Center for Interdisciplinary Exploration and Research in Astrophysics (CIERA),1800 Sherman, Evanston, IL 60201, USA}
\affiliation{NSF-Simons AI Institute for the Sky (SkAI), 172 E. Chestnut St., Chicago, IL 60611, USA}

\author[0000-0003-1749-6295]{Philipp M. Srivastava}\email{philipp.msrivastava@northwestern.edu}
\affiliation{Center for Interdisciplinary Exploration and Research in Astrophysics (CIERA),1800 Sherman, Evanston, IL 60201, USA}
\affiliation{NSF-Simons AI Institute for the Sky (SkAI), 172 E. Chestnut St., Chicago, IL 60611, USA}
\affiliation{Electrical and Computer Engineering, Northwestern University, 2145 Sheridan Road, Evanston, IL 60208, USA}

\author[0000-0002-0031-3029]{Zepei Xing}\email{Zepei.Xing1@northwestern.edu}
\affiliation{Center for Interdisciplinary Exploration and Research in Astrophysics (CIERA),1800 Sherman, Evanston, IL 60201, USA}

\begin{abstract}
Among the compact binaries detected through gravitational wave (GW) emission, double neutron stars (DNSs) hold the unique distinction of having well-characterized local analogs in the Milky Way. 
However, understanding their broader cosmological population remains challenging due to the complex metallicity-dependent evolution of their progenitors. 
Using the detailed binary population synthesis code \posydon, we model DNS formation across a cosmological range of metallicities.
We find that while DNSs are predominantly formed through common envelope evolution, metallicity-dependent radial expansion and stellar wind mass loss bifurcates the pathway into distinct subchannels, producing systematically distinct merging populations at different metallicities. 
Notably, our best representative Galactic models predict local merger rates in broad agreement with the latest GWTC-5 constraints, suggesting that the populations observed via radio pulsar surveys and through GW emission need not be in tension.
Contrary to previous studies, we demonstrate that the intrinsic merger efficiency increases toward low metallicities, a trend robust across varying binary evolution assumptions, although the magnitude and the underlying stellar processes driving this enhancement vary across metallicity regimes.
When convolved with the cosmic star formation history, our models reveal that the majority of merging DNSs originate from progenitors in the range $\sim 0.1\,Z_\odot - Z_\odot$, with significant supersolar contributions. 
Furthermore, we show that weaker stellar winds at low metallicities produce DNS mergers with systematically more symmetric mass ratios ($q \gtrsim0.9$). 
Additionally, we assess detectability within proposed configurations for third-generation GW observatories.
Our study provides a framework for interpreting the cosmological DNS population in the upcoming multi-messenger era.
\end{abstract}


\section{Introduction} \label{sec:intro}
\setcounter{footnote}{0}

Starting with the discovery of the Hulse-Taylor binary \citep{Hulse&Taylor1975}, over two dozen double neutron star (DNS)\footnote{Throughout this work, we use DNS to refer to the general astrophysical population, while binary neutron star (BNS) or merging DNS refers to the subset observable via gravitational waves.} candidates in the Galaxy have now been identified, roughly half of which will merge within a Hubble time due to GW radiation \citep{Tauris+2017}. 
Since radio pulsar surveys primarily target the Galactic plane, most of these observed systems reside in the Milky Way disk and are associated with relatively recent star formation (with most systems born within the past $\sim 1 \, \rm{Gyr}$), when the progenitor metallicities are expected to be near-solar \citep[e.g.,][]{Snaith+2015, Bland-Hawthorn&Gerhard2016, Xiang&Rix2022}.
As a result, our current understanding of DNS formation is calibrated exclusively to the local Galactic conditions, making the interpretation of low-metallicity observations reliant on theoretical extrapolation.

In the era of multi-messenger astrophysics, this limitation can be profound. Third generation (3G) observatories like Cosmic Explorer and the Einstein Telescope \citep[][]{Dwyer+2015, Abbott+2017, Chamberlain&Yunes2017, Reitze+2019, Kalogera+2021, Abac_ET+2026} are projected to actively detect BNS mergers originating from across cosmic time \citep[probing redshifts beyond cosmic noon;][]{Evans+2017, Maggiore+2020}. Furthermore, time-domain surveys such as the Nancy Grace Roman Space Telescope \citep{Spergel+2015} and the Vera C. Rubin Observatory \citep{2019ApJ...873..111I} will soon sample kilonovae and gamma-ray burst (GRB) afterglows born from BNS mergers across diverse star-forming environments spanning a wide range of metallicities. Interpreting these observations requires DNS models informed by the Galactic observations yet extrapolated self-consistently to other metallicities, and a way to test whether such Galactic calibrations hold beyond the Milky Way.

Since the pioneering works by \citet{Flannery&vandenHeuvel1975, Bhattacharya&vandenHeuvel1991, Tutukov&Yungelson1993, Lipunov+1997, Zwart&Yungelson1998}, multiple studies have modeled DNS formation from isolated massive binaries using rapid population synthesis codes \citep[e.g.,][]{Belczynski+2002, Dominik+2012, Kruckow+2018, Vigna-Gomez+2018, Chattopadhyay+2020, Sgalletta+2023}, including examining its dependence on metallicity \citep[e.g.,][]{Chruslinska+2018, Giacobbo&Mapelli2018, Mapelli&Giacobbo2018, Neijssel+2019, Mapelli+2019, Santoliquido+2021, broekgaarden+2022, Iorio+2023, SKRoy+2025, Pellouin+2025, Venet&Dvorkin2026}.
These studies generally find that BNS formation is largely metallicity-independent \citep[see][and references therein]{vanSon+2025}.  While rapid population synthesis codes allow for computationally inexpensive simulations, the binary modeling is approximated by analytical fits to single stellar evolutionary tracks \citep{Hurley+2000, Hurley+2002}. Consequently, the evolution of stars driven out of thermal equilibrium, such as during mass transfer phases, may not be modeled accurately; effects that can depend on metallicity \citep[e.g.,][]{Klencki+2020} and compound over the $\mathcal{O}(10 \, \rm Myr)$ lifetime of massive stars. In contrast, detailed population synthesis codes like \texttt{BPASS} \citep{Eldridge+2017} and \posydon \citep{Fragos+2023, Andrews+2025} overcome some of these limitations. In particular, \posydon self-consistently models binary evolution using state-of-the-art stellar evolution grids generated using \mesa \citep{Paxton+2011, Paxton+2013, Paxton+2015, Paxton+2018, Paxton+2019,Jermyn+2023}\footnote{Even more self-consistent approaches following both stars continuously through successive interaction phases are currently limited to
specific environments \citep[e.g.,][]{Xu+2025}}. 

Additionally, \posydon provides the internal profiles of the donor stars at the onset of the mass transfer, from which the envelope binding energy and the core-envelope boundary can be determined. While DNS formation is primarily driven by common envelope (CE) evolution \citep[see][and references therein]{Broekgaarden+2026}, a recent study using the detailed solar metallicity grids in \posydon \citep{Chattaraj+2026a} showed that depending on the evolutionary state of the donor at CE onset, this dominant formation channel bifurcates into two distinct subchannels: systems where the donor possesses a helium core (Case B mass transfer, or \Hecore channel) and those with a more evolved carbon-oxygen core (Case C, or \COcore channel). These subchannels leave imprints on the orbital period distributions of the resulting DNSs. Indeed, the observed orbital periods of Galactic DNSs span over three orders-of-magnitude, and \citet{Andrews&Mandel2019} found
evidence that the systems cluster into distinct subpopulations. In \citet{Chattaraj+2026a}, we
demonstrated that the \Hecore and \COcore subchannels can reproduce this clustering, highlighting the importance of using detailed binary evolution models capable of resolving the stellar structures and profiles at the onset of mass transfer.

In this work, we investigate the metallicity-dependent formation and evolution of DNSs using the \posydon code. 
In Section \ref{sec:methods}, we describe the simulation setup for this study. Section \ref{sec:form_channels} explores the metallicity-dependent stellar and binary evolution that affects DNS formation.
In Section \ref{sec:efficiency}, we discuss the intrinsic merger efficiencies as a function of metallicity and break down the cosmological population by its progenitor metallicities. Section \ref{sec:merger_rates} presents the redshift evolution of the merger rate density for our model populations. We explore the BNS mass distributions in Section \ref{sec:masses_and_detectability}, and discuss prospects for detection with next-generation GW observatories in Section \ref{sec:detectability}. Finally, Section \ref{sec:conclusions} presents the conclusions of this study and discusses its limitations.

\section{Methods} \label{sec:methods}

\setlength{\tabcolsep}{8.2pt}
\begin{table*}[]
\centering
\caption{Summary of the model populations and their key binary physics assumptions. The columns list the model name, the energy conversion efficiency for common envelope ejection $\alpha_{\rm CE}$, the adopted hydrogen fraction defining the core-envelope boundary $X_{\rm core}$, the supernova model, and the natal kick prescription. To test the metallicity preference of BNS formation, we report the relative merger efficiency for each population model (per unit star-forming mass) between our lowest metallicity $10^{-4}\Zsun$ and solar metallicity, which is further broken into two regimes, from $\Zsun$ to $0.1\Zsun$, and $0.1\Zsun$ to $10^{-4}\Zsun$ (Section \ref{sec:efficiency}). For each model population, we also provide the cumulative annual detection rate out to a redshift $z=2$ for the next-generation Cosmic Explorer ($40\,\mathrm{km} + 20\,\mathrm{km}$) and Einstein Telescope (2L $15\,\mathrm{km}$) detector networks (Section \ref{sec:detectability}). Models shown in boldface indicate the ``observationally calibrated" subset that best reproduces the observed DNSs in the Milky Way (Section \ref{subsec:galactic_dns}).}
\begin{tabular}{ccccccccccc}
\hline
\hline
\multirow{2}{*}{Model} & \multicolumn{2}{c}{$\textrm{CE parameters}$} & \multirow{2}{*}{SN\tablenotemark{a}} & \multirow{2}{*}{Kick\tablenotemark{b}}
& \multicolumn{3}{c}{\textrm{Efficiency Ratio}} & & \multicolumn{2}{c}{$\mathcal{N}\,(z \leq 2)$ $[10^3\,\mathrm{yr}^{-1}]$} \\
\cline{2-3}
\cline{6-8}
\cline{10-11}
\noalign{\smallskip}
& $\mathrm{\alpha_{CE}}$ & $X_{\rm core}$ & & & $\frac{\eta\,(10^{-4}\Zsun)}{\eta \, (0.1\Zsun)}$ & $\frac{\eta\,(0.1\Zsun)}{\eta\,(\Zsun)}$ & $\frac{\eta\,(10^{-4}\Zsun)}{\eta\,(\Zsun)}$ & & CE & ET \\
\noalign{\smallskip}
\hline
\noalign{\smallskip}
\texttt{MODEL00} & 1 & 30\% & PS20 & AsymEj & $\sim$ & 6.5 $\uparrow$ & 5.7 $\uparrow$ & & 32.1 & 20.8 \\
\texttt{MODEL01} & 1 & 30\% & PS20 & lognormal & 0.5 $\downarrow$ & 3.0 $\uparrow$ & 1.5 $\uparrow$ & & 3.2 & 2.0\\
\texttt{MODEL02} & 1 & 30\% & PS20 & Maxwellian & 0.6 $\downarrow$ & 1.6 $\uparrow$ & $\sim$ & & 1.5 & 1.0\\
\pmb{\texttt{MODEL03}} & 1 & 30\% & S16 & AsymEj & 1.5 $\uparrow$ & 6.7 $\uparrow$ & 9.8 $\uparrow$ & & 52.9 & 34.5\\
\texttt{MODEL04} & 1 & 30\% & S16 & lognormal & $\sim$ & 2.5 $\uparrow$ & 2.6 $\uparrow$ & & 4.1 & 2.6\\
\texttt{MODEL05} & 1 & 30\% & S16 & Maxwellian & 0.4 $\downarrow$ & 2.8 $\uparrow$ & 1.2 $\uparrow$ & & 1.5 & 1.0\\
\texttt{MODEL06} & 1 & 30\% & F12d & AsymEj & $\sim$ & 1.8 $\uparrow$ & 1.7 $\uparrow$ & & 118.6 & 77.2\\
\texttt{MODEL07} & 1 & 30\% & F12d & lognormal & $\sim$ & 1.8 $\uparrow$ & 1.7 $\uparrow$ & & 7.3 & 4.7\\
\texttt{MODEL08} & 1 & 30\% & F12d & Maxwellian & 0.8 $\downarrow$ & 1.6 $\uparrow$ & 1.3 $\uparrow$ & & 2.2 & 1.4\\
\pmb{\texttt{MODEL09}}\tablenotemark{*} & 1 & 30\% & S16 & AsymEj & 1.4 $\uparrow$ & 6.8 $\uparrow$ & 9.6 $\uparrow$ & & 47.9 & 31.2\\
\pmb{\texttt{MODEL10}}\tablenotemark{\#} & 1 & 30\% & S16 & AsymEj & 1.3 $\uparrow$ & 2.2 $\uparrow$ & 2.8 $\uparrow$ & & 100.1 & 64.4\\
\pmb{\texttt{MODEL11}} & 3 & 10\% & S16 & AsymEj & 6.4 $\uparrow$ & 4.1 $\uparrow$ & 26.0 $\uparrow$ & & 108.5 & 67.6\\
\texttt{MODEL12} & 3 & 10\% & S16 & lognormal & 4.3 $\uparrow$ & 5.0 $\uparrow$ & 21.4 $\uparrow$ & & 11.1 & 7.0\\
\texttt{MODEL13} & 3 & 10\% & S16 & Maxwellian & 3.0 $\uparrow$ & 4.8 $\uparrow$ & 14.6 $\uparrow$ & & 3.6 & 2.2\\
\pmb{\texttt{MODEL14}} & 3 & 10\% & S16 & Linear & 8.6 $\uparrow$ & 9.0 $\uparrow$ & 77.5 $\uparrow$ & & 20.9 & 13.0\\
\texttt{MODEL15} & 0.5 & 30\% & S16 & AsymEj & 0.5 $\downarrow$ & 5.0 $\uparrow$ & 2.5 $\uparrow$ & & 46.8 & 30.6\\
\texttt{MODEL16}\tablenotemark{\#} & 0.5 & 30\% & S16 & AsymEj & 0.5 $\downarrow$ & 3.1 $\uparrow$ & 1.4 $\uparrow$ & & 75.0 & 48.1\\
\texttt{MODEL17}\tablenotemark{\#} & 2 & 30\% & S16 & AsymEj & 3.7 $\uparrow$ & 1.2 $\uparrow$ & 4.4 $\uparrow$ & & 221.6 & 141.9\\
\texttt{MODEL18}\tablenotemark{\#} & 1 & 30\% & S16 & lognormal & 0.9 $\downarrow$ & 1.8 $\uparrow$ & 1.6 $\uparrow$  & & 6.1 & 4.0\\
\noalign{\smallskip}
\hline 
\hline
\end{tabular}
\begin{flushleft}
\tablenotetext{}{$\uparrow\,: \eta$ goes up towards lower metallicity; $\downarrow\,: \eta$ goes down towards lower metallicity; $\sim\,: \eta$ remains within 10\% of the bounds.}
\tablenotetext{a}{S16: \citet{Sukhbold+2016}; PS20: \citet{Patton&Sukhbold2020}; F12d: \cite{Fryer+2012}$-$delayed. For both S16 \& PS20, we use the N20 engine.}
\tablenotetext{b}{AsymEj: \citet{Janka2017} with exact scaling in Eq. 2 of \citet{Chattaraj+2026a}; lognormal: \citet{Disberg&Mandel2025}; Maxwellian: \citet{Hobbs+2005} with $\sigma_{\rm CCSN}=265\,\rm{km/s}$ and $\sigma_{\rm ECSN}=20\,\rm{km/s}$; Linear: \citet{Richards+2023}. 
The ``AsymEj" and ``Linear" models constitute our ``physics based" kick prescriptions referenced throughout this work.}
\tablenotetext{*}{The initial binary fraction follows the mass-dependent prescription from \citet{Moe&DiStefano2017}.} \tablenotetext{\#}{ Low He star wind version.}
\end{flushleft}
\label{tab:model_populations}
\end{table*}

\subsection{Simulation Setup}
\label{subsec:setup}
We use the detailed binary population synthesis code \posydon \footnote{The version of the code is identified by the commit hash \texttt{888d89cc5} available at \href{https://github.com/POSYDON-code/POSYDON/}{https://github.com/POSYDON-code/POSYDON/}. The \posydon v2.3 datasets are available on Zenodo under an open-source Creative Commons Attribution license here: \dataset[doi: 10.5281/zenodo.21909570]{https://zenodo.org/records/21909570}.} \citep{Fragos+2023, Andrews+2025} to model synthetic populations of binaries at eight metallicities: [$10^{-4}, 10^{-3}, 10^{-2}, 0.1, 0.2, 0.45, 1, 2$] $\Zsun$, with $\Zsun=0.0142$ \citep{Asplund+2009}. \posydon relies on pre-computed grids of detailed single and binary star models generated using the 1D stellar evolution code \mesa \citep{Paxton+2011, Paxton+2013, Paxton+2015, Paxton+2018, Paxton+2019,Jermyn+2023}. The evolution of detached, non-interacting binaries are modeled using grids of single H-rich and He-rich stars (\texttt{single-HMS} and \texttt{single-HeMS}, respectively). For interacting binaries, mass transfer is modeled self-consistently with the evolution of the orbit through three binary star grids, namely the \texttt{HMS-HMS} grid (for two H-rich stars), the \texttt{CO-HMS} grid (for a compact object and a H-rich star), and the \texttt{CO-HeMS} grid (for a compact object and a He-rich star representing a donor stripped of its H envelope). At each metallicity, we sample $10^6$ binaries using a \citet{Kroupa2001} initial mass function (IMF) for the primary star, and a secondary mass from a distribution that is flat in mass ratio. For implementations pertaining to metallicity-dependent stellar and binary evolution, we refer the reader to \citet{Fragos+2023} and \citet{Andrews+2025}, while the setups specific to DNS formation are identical to \citet{Chattaraj+2026a}. Finally, to construct cosmologically respresentative populations, we weigh our simulated binaries with the star formation history (SFH) and metallicity evolution of the Universe \citep[see Section 6.2 in][]{Andrews+2025} using the large-scale cosmological simulation \texttt{IllustrisTNG} \citep{Marinacci+18, Pillepich+18, Springel+18, Nelson+18, Naiman+18} and the empirical model of \citealt{Madau&Fragos2017} (hereafter, \texttt{MF17}). The complete set of population models\footnote{This dataset will be released upon publication of this work.} considered in this work is presented in Table \ref{tab:model_populations}. We systematically explore uncertainties in binary evolution, with the models varying the common envelope efficiency ($\alpha_{\mathrm{CE}} \sim 0.5$--$3$), the hydrogen mass fraction defining the core-envelope boundary ($X_{\mathrm{core}} \in [0.1, 0.3]$), the core-collapse remnant mass prescriptions, and the natal kick models (see the notes to Table \ref{tab:model_populations} for details). 

\subsection{Modifications to our grid architecture}
\label{subsec:setup_modifications}

In addition to the standard \posydon datasets described above, we introduce two modifications necessary for this work. First, following the approach at solar metallicity in \citet{Chattaraj+2026a}, we supplement our \texttt{CO-HMS} grid across all eight metallicities with densely sampled sets of models, adding a total of 23,550 \mesa models (Table \ref{tab:dense_CO-HMS_RLO_grid}) to the baseline 13,464 models per metallicity \citep[Table 1 in][]{Andrews+2025}.\footnote{The supplementary grid consisting of high-resolution models combined with the \posydon v2.0 grids are available on Zenodo under data release v2.3: \dataset[doi: 10.5281/zenodo.21909570]{https://zenodo.org/records/21909570}.} These supplementary grids provide increased resolution to the binary tracks relevant for DNS formation. Moreover, the range of the parameter space shifts with each metallicity grid (see Table \ref{tab:dense_CO-HMS_RLO_grid}), enabling accurate modeling of the metallicity-dependent evolutionary pathways we find. In addition to the current work, these supplementary grids have previously been used to construct the multi-metallicity DNS populations presented in \citet{Chattaraj+2026b}.

Second, in a previous study \citep{Chattaraj+2026b}, we demonstrate that the standard \posydon implementation of extrapolating the Wolf-Rayet wind prescription of \citet{Nugis&Lamers2000} into the regime of low-mass, low-luminosity stripped stars can significantly overestimate mass loss rates compared to recent observations \citep{Gotberg+2023, Drout+2023}. This enhanced mass loss drives increased orbital expansion prior to the second supernova (SN), with the effect becoming more pronounced at high metallicities \citep[see Figure 8 in][]{Chattaraj+2026b}.
While the Monte Carlo radiative transfer models in \citet{Vink2017} may offer a more accurate representation of observations \citep{Ramachandran+2024}, the true mass loss rates may be even lower \citep{Gotberg+2023}. Implementing a new mass loss prescription within \posydon would require regenerating the detailed grids from scratch, a task we defer to future iterations of the code.

For this work, we adopt the mitigation technique introduced in \citet{Chattaraj+2026b}: we simulate a set of population models (marked by $\#$ in Table \ref{tab:model_populations}) where the evolution of He star$-$NS systems (represented by the \texttt{CO-HeMS} grid) at all metallicities is artificially set to follow the corresponding evolutionary track of an equivalent binary at $10^{-4}\,Z_\odot$ metallicity \citep[for more details, see Appendix in][]{Chattaraj+2026b}. This treatment effectively limits excessive orbital expansion due to overestimation of the stripped star winds. For detailed \posydon studies of the evolution and eventual collapse of stripped stars, we refer the reader to \citealt{Souropanis+2026}, \citealt{Zapartas+2026} and \citealt{Briel+2026c} \citep[see also][]{Aguilera-Dena+2022}.

\subsection{Constraints from the Galactic DNS sample}
\label{subsec:galactic_dns}

While our population synthesis explores a broad range of binary evolution assumptions, we use the observed Galactic DNS population to identify a physically representative subset of models. In our preceding study \citep{Chattaraj+2026a}, we demonstrated that our solar-metallicity models reproduce the orbital period--eccentricity ($P_{\rm orb}-e$) distribution of the observed sample of 25 Galactic DNS systems, with the exception of three systems at very short orbital periods and $e\sim0.6$, which may have a dynamical origin (\citealt{Andrews&Mandel2019}, although see \citealt{Grichener+2026}). The same models were also consistent with the BNS merger-rate constraints from GWTC-3. While we didn't perform a rigorous Bayesian statistical comparison due to the lack of a detailed pulsar evolution model within \posydon, we were nonetheless able to impose stringent constraints on massive binary evolution leading to DNS formation, which we summarize below.

Specifically, \citet{Chattaraj+2026a} found that reproducing the observations imposed three broad constraints.
First, the preferred models adopt ``physics-based'' natal-kick prescriptions (``AsymEj" and ``Linear" kick models from Table \ref{tab:model_populations}). Second, among the SN prescriptions explored, the \citet{Sukhbold+2016} model produces the best agreement with the Galactic sample. Third, reproducing the over three order-of-magnitude range of observed orbital periods requires both the \Hecore and \COcore formation channels, which requires either a core-envelope boundary definition of $X_{\rm H}\simeq30\%$ or a sufficiently high CE ejection efficiency, $\alpha_{\rm CE}\gtrsim1.2$ (see Fig. 10 of \citealt{Chattaraj+2026a}).

Informed by that study, we designate models satisfying these criteria as our ``observationally calibrated'' subset and use them as an empirical baseline throughout this work. These models are highlighted in boldface in Table \ref{tab:model_populations}. We explicitly emphasize this set of models throughout the figures in this work to assess how a binary population consistent with the Galactic DNS observations evolves across metallicity.

\section{Metallicity dependence of Double Neutron Star formation}
\label{sec:form_channels}

Several previous studies have highlighted the critical role of the CE phase leading to DNS formation \citep[see][and references therein]{Broekgaarden+2026}. Recently, we demonstrated the importance of using detailed stellar evolution models to resolve the internal structure of the CE donor in \citet{Chattaraj+2026a}. Using \posydon, that study identified a bifurcation in the CE channel at solar metallicity, driven by the evolutionary state of the donor star at the onset of RLO. The two channels include the \Hecore channel (Case B mass transfer), where the donor fills its Roche lobe as a less evolved giant star with a helium-rich core (burning hydrogen in a shell, or in the rare case, undergoing core helium burning), and the \COcore channel (Case C mass transfer), with a highly evolved donor star with a carbon-oxygen core. Together, the two channels are able to broadly explain the observed DNS subpopulations in the Milky Way Field (however, see \citealt{Grichener+2026}; Fig. 3 in \citealt{Valli+2025}). The characteristics of these two channels is expected to depend on metallicity, which modifies both the internal structure of the H-rich donor and the evolution of the binary orbit prior to the CE phase. We first consider the effect of metallicity on the H-rich donor star before examining its consequences for the binary orbit.

\subsection{Metallicity dependent single stellar physics}
\label{subsec:single_star}

\subsubsection{Radial expansion of the H-rich star}
\label{subsubsec:radius_evol}

\begin{figure}
    \centering
    \includegraphics[width=1.00\linewidth]{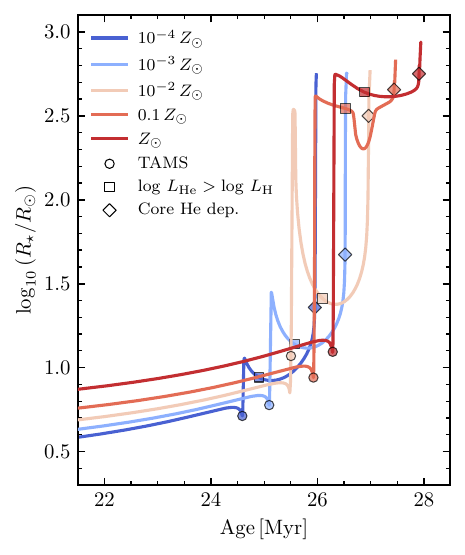}
    \caption{The radial expansion of a $\simeq 10.2\Msun$ H-rich star at five representative metallicities. At lower metallicities, the star remains relatively compact during the Hertzsprung Gap (HG) phase, delaying significant expansion until advanced evolutionary stages. Consequently, donors in this regime are highly evolved at the onset of RLO. On the other hand, higher metallicity models experience rapid HG expansion immediately after terminal age main sequence (TAMS), allowing donors to initiate RLO early in their post-main-sequence evolution as well as during the more evolved giant branches.}
    \label{fig:radius_evol}
\end{figure}

\begin{figure*}
    \centering
    \includegraphics[width=0.95\linewidth]{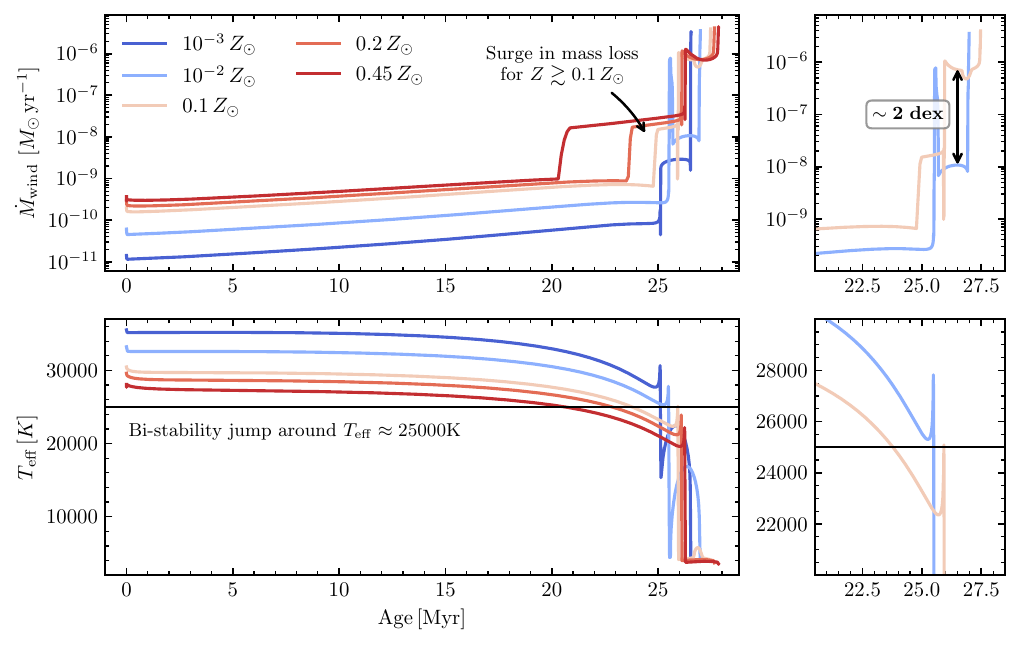}
    \caption{Evolution of the wind mass-loss rate (top panel) and effective temperature (bottom panel) for a $\simeq 10.2\,\Msun$ H-rich model across five initial metallicities. The horizontal black line in the bottom panel (and its inset) denotes the bi-stability jump at $T_{\mathrm{eff}} \approx 25,000\,\mathrm{K}$ (see Section \ref{subsubsec:wind_mass_loss}). The inset panels isolate the late-stage, post-main-sequence evolution of the $Z=10^{-2}\Zsun$ and $0.1\Zsun$ models. As high-metallicity stars ($Z \gtrsim 0.1\,\Zsun$) expand and cool on the main sequence, they cross the bi-stability jump and experience a surge in the wind mass loss rate. Furthermore, as highlighted in the top inset, there is a ${\sim}2\,\mathrm{dex}$ enhancement in the supergiant wind mass-loss rate for $Z \gtrsim 0.1\Zsun$ compared to metal-poor models, which remain hotter and more compact. Ultimately, this extreme mass loss forces significant orbital widening in these high-metallicity binaries prior to RLO, leading to a distinct outcropping of models in our \texttt{CO-HMS} grid (Figure \ref{fig:CO-HMS_RLO_multi_panel}), responsible for the bifurcation within the CE formation channel.}
    \label{fig:bi-stability_jump}
\end{figure*}

Higher initial metallicity increases the mean molecular weight of a star, resulting in larger stellar radii during the main sequence. During later phases, metallicity enhances its radiative opacity and drives radial expansion \citep[e.g.,][]{Sanyal+2017, Xin+2022}, thereby determining the evolutionary stage at which a donor fills its Roche lobe (for a given initial orbital period). In Figure \ref{fig:radius_evol}, we show the radial evolution of a $\simeq 10.2\Msun$ H-rich star, representative of the mass range for NS progenitors in our models, at five metallicities. 

At low metallicities (typically $\lesssim 10^{-2}\,Z_\odot$), the star remains compact during the Hertzsprung Gap (HG; between TAMS and $\log L_{\rm He} > \log L_{\rm H}$ in Fig. \ref{fig:radius_evol}) phase, extending to radii $\lesssim 100\Rsun$. This suppressed expansion is due to a combination of reduced envelope opacity and elevated central temperatures following core-H exhaustion \citep[][]{Klencki+2020}. In this metallicity regime, most donors do not reach the radii required to overfill their Roche lobes until well after core-He depletion, leading to Case C mass transfer scenarios. 

In contrast, higher metallicity ($\gtrsim 0.1\Zsun$) donors experience significant radial expansion during the HG phase. Thus, depending upon the binary separation, these donors fill their Roche lobes either during the HG phase or later during subsequent expansion following core-He depletion, enabling both Case B and Case C RLO. Case B donors, being less evolved and more compact, possess higher envelope binding energies than their Case C counterparts \citep[see Fig. 10 in][]{Chattaraj+2026a}. Consequently, ejecting a Case B envelope during the CE phase requires a greater expenditure of orbital energy, making them less likely to survive the interaction. However, systems that do survive have tighter post-CE orbits than their Case C counterparts. As shown in \citet{Chattaraj+2026a}, at solar metallicity, these tighter Case B systems constitute the population that are able to merge within a Hubble time. However, this paradigm shifts at lower metallicities (see Sections \ref{subsec:CO-HMS_RLO} and \ref{subsec:all_form_channels}). 

\subsubsection{Stellar wind mass loss of the H-rich star}
\label{subsubsec:wind_mass_loss}

In addition to affecting the stellar radii, metallicity also modulates the wind mass-loss rate of the donor star, driving orbital expansion and envelope mass loss prior to the CE phase. In Figure \ref{fig:bi-stability_jump}, we show the wind mass loss rate and the effective temperature evolution of the same $\simeq 10.2\Msun$ H-rich star as in Figure~\ref{fig:radius_evol}, at the same five representative metallicities. Two related metallicity-dependent effects are apparent.

First, as high-metallicity stars ($\gtrsim 0.1\,Z_\odot$) expand and cool on the main sequence, their effective temperatures drop below $T_{\mathrm{eff}} \approx 25,000\,\mathrm{K}$ (bottom panel). The \texttt{Dutch} wind prescription in \mesa adopted in our \posydon grids then triggers the bi-stability jump \citep{Lamers+1995, Vink+1999} driven by iron recombination (\ion{Fe}{4} $ \rightarrow$ \ion{Fe}{3}).The resulting increase in the opacity leads to a surge in the mass-loss rate (top panel of Figure~\ref{fig:bi-stability_jump}).\footnote{We note that the existence and strength of the bi-stability jump are debated. Recent observational analyses of Galactic and LMC OB-stars find little evidence for a sharp jump \citep{deBurgos+2024, Verhamme+2024}, while other theoretical studies support it
\citep{Petrov+2016, Keszthelyi+2017, Krticka+2024, Vink+2026}. Nevertheless, the uncertainty in mass loss rates in this regime does not impact the conclusions of this paper (Section \ref{subsubsec:wind_mass_loss}).} In contrast, the lower metallicity ($\lesssim10^{-2}\Zsun$) models remain hotter and more compact prior to evolving into giants, avoiding the bi-stability jump. Although, we note here that despite the surge in wind mass loss associated with the bi-stability jump, its integrated impact on the binary orbit is negligible compared to the subsequent mass loss during the supergiant phase, even at high metallicity.

Second, and more importantly, the higher metallicity models exhibit extreme supergiant winds following the \citet{deJager+1988} prescription, which scales with the star's effective temperature and luminosity, effectively translating to a steep radius dependence at fixed luminosities. At high metallicities ($\gtrsim 0.1\Zsun$), the donors tend to acquire the large radii and lower temperatures characteristic of red or yellow supergiants, for which the \citet{deJager+1988} prescription predicts significantly stronger mass-loss rates, than for the hotter low-metallicity models, which evolve into more compact blue supergiants \citep{Ou+2023}.
These differences are large enough to cause a two order-of-magnitude difference in the wind mass-loss rate (top inset panel of Figure \ref{fig:bi-stability_jump}). As a consequence, strong supergiant winds at $\gtrsim 0.1\Zsun$ forces significant orbital widening, delaying the onset of RLO in addition to reducing the donor mass.

\subsection{Metallicity dependent evolution of the H-rich star$-$NS binary}
\label{subsec:CO-HMS_RLO}

\begin{figure*}
    \centering
    \includegraphics[width=1.00\linewidth]{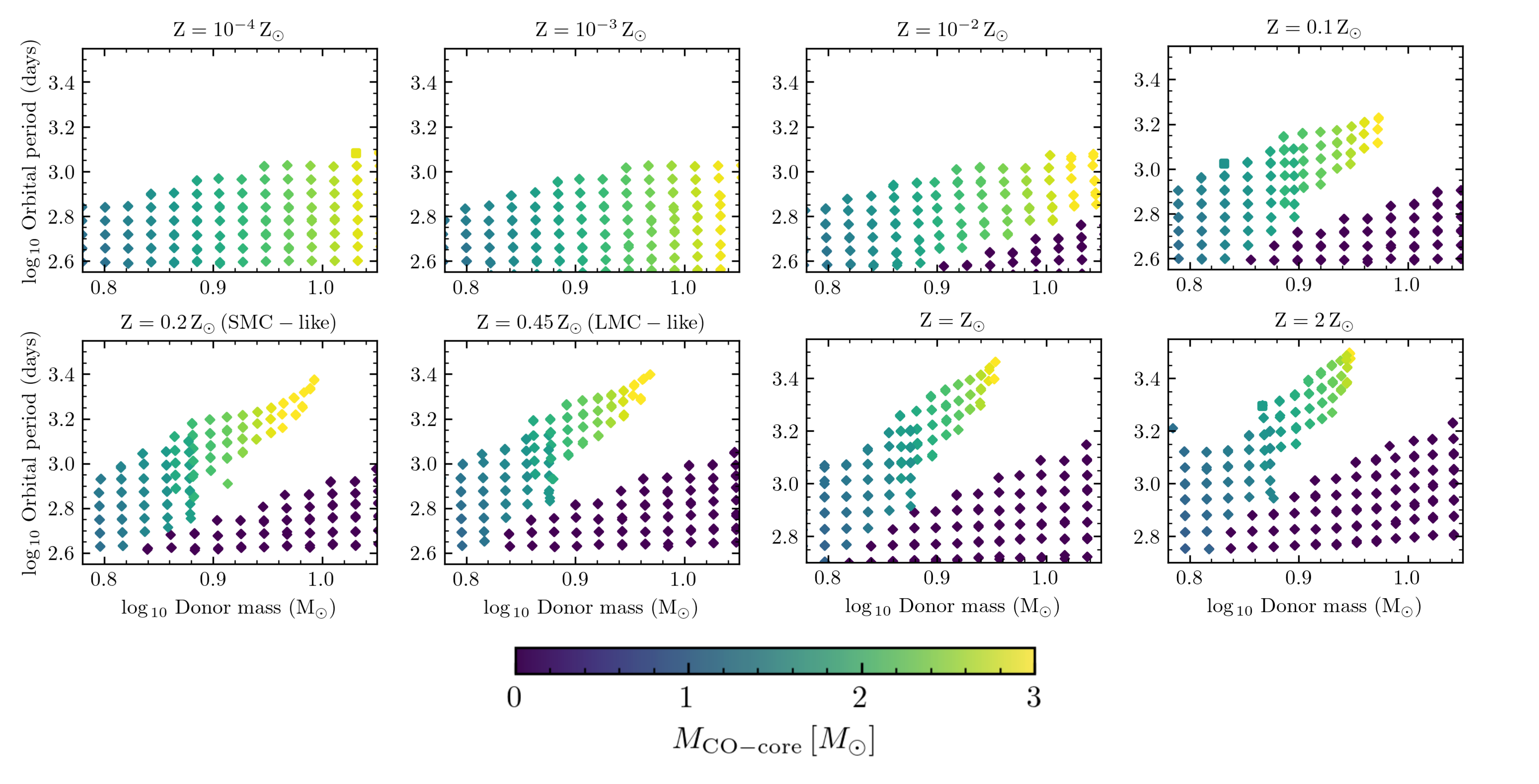}
    \caption{Slices of the high resolution supplementary set of \texttt{CO-HMS} models terminating in unstable RLO across eight metallicities. The color coding denotes the donor's carbon-oxygen core mass at the onset of RLO. At low metallicities ($Z \lesssim 10^{-2}\,Z_\odot$), suppressed radial expansion during the HG phase ensures that all donors are highly evolved with developed C/O cores (Section \ref{subsubsec:radius_evol}). At higher metallicities ($Z \gtrsim 0.1\,Z_\odot$), two distinct physical effects drive a bifurcation in the parameter space. First, enhanced radial expansion allows less-evolved donors to fill their Roche lobes before developing a C/O core (dark purple models, $M_{\mathrm{CO-core}} = 0$), leading to DNS formation through the \Hecore formation channel. Second, extreme supergiant wind mass loss strips the envelopes of the highly evolved donors, forcing significant orbital widening prior to RLO (Section \ref{subsubsec:wind_mass_loss}). This supergiant wind-driven evolution shifts the corresponding \COcore channel into a distinct upper-left outcropping characterized by longer orbital periods and lower donor masses (Section \ref{subsec:CO-HMS_RLO}). Ultimately, the orbital expansion is so severe for $Z \gtrsim \Zsun$ (note the different y-axis ranges for these panels) that the \COcore channel fails to produce any BNSs that merge (Section \ref{subsec:all_form_channels}).
    }
    \label{fig:CO-HMS_RLO_multi_panel}
\end{figure*}

\begin{figure*}
    \centering
    \includegraphics[width=1.00\linewidth]{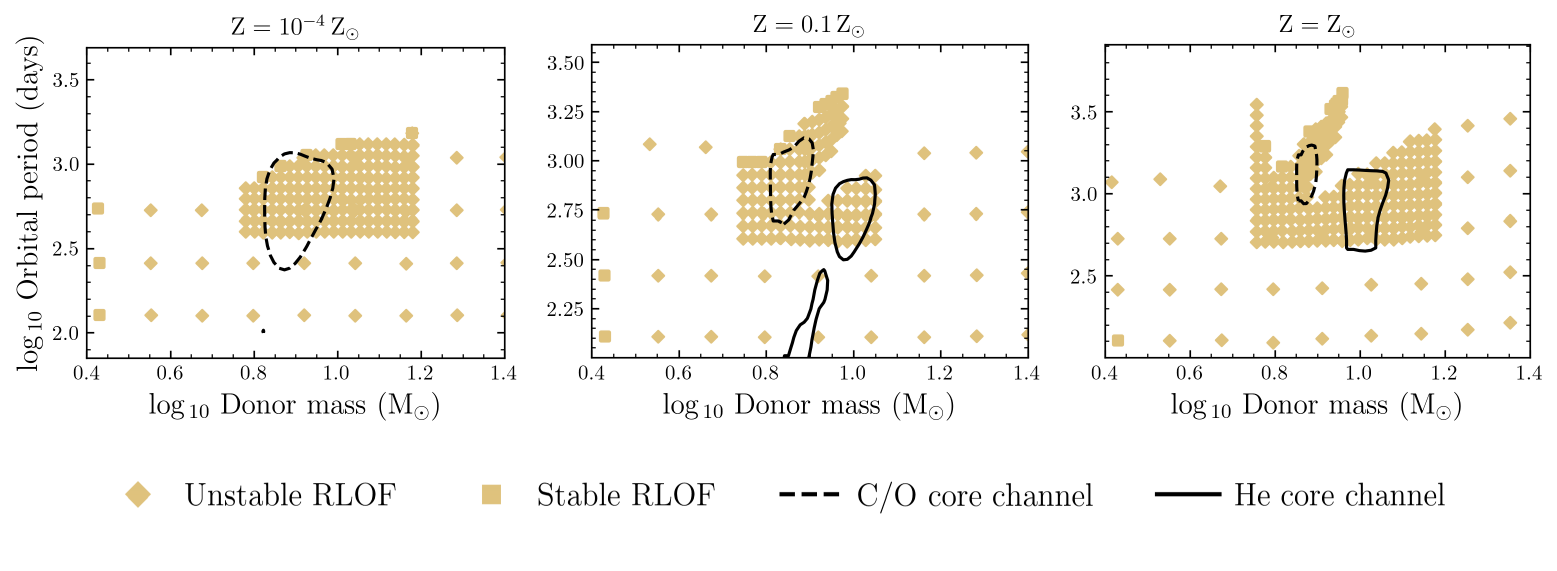}
    \caption{Slices of the \texttt{CO-HMS\_RLO} grids at $Z = 10^{-4}\,Z_\odot$ (left), $0.1\,Z_\odot$ (middle), and $Z_\odot$ (right), for a $1.43\,M_\odot$ NS and a H-rich donor at the onset of RLO. The grids combine the standard \posydon DR2 grids with our supplementary models (Section~\ref{subsec:setup_modifications}). Diamond markers denote systems that undergo CE evolution and square markers represent stable RLO. Contours show the 75th-percentile distributions of the \Hecore (solid) and \COcore (dashed) subchannels in \texttt{MODEL03}.
    }
    \label{fig:CO-HMS_RLO_populations}
\end{figure*}

We now examine how the combined effects of metallicity on the donor's radial expansion and wind mass loss determine the evolution of the binary prior to the CE phase. Figure \ref{fig:CO-HMS_RLO_multi_panel} shows slices of the supplementary high-resolution \texttt{CO-HMS\_RLO} grid slices at the eight metallicities considered in this work (Table \ref{tab:dense_CO-HMS_RLO_grid}). This grid follows the evolution of a NS$-$H-rich star through RLO, with each model started at a point on a regular grid of NS mass, H-rich donor mass, and initial orbital period. Most models terminate in an unstable RLO phase\footnote{See Section 4.2.4 of \citet{Fragos+2023} for the criteria used to classify unstable RLO within \posydon.} of the H-rich star$-$NS binary (diamond markers; while square markers denote rare cases of stable RLO), and the color scale indicates the C/O-core mass of the donor at the onset of the CE phase. Binaries that do not undergo mass transfer are omitted from this plot (white space).

A transition emerges at $\simeq10^{-2}\Zsun$. At lower metallicities, the suppressed HG radial expansion implies that donors are all evolved stars with a massive C/O core (see the color bar) at the onset of RLO (this also applies to interacting systems outside the range of this high-resolution grid). DNS formation therefore proceeds almost entirely via the \COcore channel. At higher metallicities, the donor star can initiate a CE while on its HG expansion phase, when it has a He-rich core (see the models with zero C/O core mass), in addition to interacting at later evolutionary stages. The \Hecore and \COcore channels therefore coexist at metallicities $>10^{-2}\Zsun$.

Furthermore, the increased mass-loss-driven orbital expansion shapes the structure of the models at high metallicity. Due to the strong supergiant winds (Section \ref{subsubsec:wind_mass_loss}), the donor loses substantial mass prior to initiating CE, while simultaneously expanding the orbit. This shifts the interacting systems toward lower donor masses (leftward) and longer orbital periods (upwards), producing the characteristic extended ``peninsula”-like outcropping in the $P_{\mathrm{orb}}-M_{\mathrm{donor}}$ plane (see Figure \ref{fig:CO-HMS_RLO_multi_panel} for $Z\geq0.1\Zsun$). A detailed discussion of the binary evolution models surrounding the ``peninsula"-like outcropping at solar metallicity is given in Fig. 6 and Sec. 3.2 of \citet{Chattaraj+2026a}. 

We draw the conclusion from these detailed evolutionary tracks that while at low metallicities ($\lesssim10^{-2}Z_\odot$), the CE phase proceeds exclusively through the \COcore channel, at higher metallicities, the evolution bifurcates as a result of metallicity-dependent stellar and binary physics, opening up both the \Hecore and \COcore channels, with the latter existing at increased orbital periods and lower donor masses.

\subsection{Metallicity dependent DNS formation from isolated binary evolution} 
\label{subsec:all_form_channels} 

Compared to other compact binaries with multiple formation pathways \citep{Belczynski+2016, deMink&Mandel2016, Marchant+2016, Giacobbo+2018, Bavera+2021, Zevin+2021, Gallegos-Garcia+2021, vanSon+2022, Xing+2024, Maclean+2026, Briel+2026a, Gallegos-Garcia+2026}, DNS formation via isolated binary evolution is simplified, heavily dominated by the CE channel \citep{Broekgaarden+2026}. In \citet{Chattaraj+2026a}, we demonstrate that the CE channel accounts for $\gtrsim 95\%$ of all DNS formation at solar metallicity. Although at solar metallicity, up to $\sim 5\%$ of DNS systems can form through a non-interacting/stable RLO channel, we find this fraction decreases toward lower metallicities. In our lowest metallicity populations ($10^{-4}\,Z_\odot$), all DNS formation involves a CE phase. Overall, we find that the dominance of the CE channel persists across the metallicity range explored here.

In Section \ref{subsec:CO-HMS_RLO}, we discussed the bifurcation of the CE channel based on the type of donor star. Here, we shall extend this framework to distinguish between merging and non-merging DNS systems:

\begin{itemize}
    \item $Z\lesssim 10^{-2}\Zsun$: With delayed radial expansion, donors tend to only overfill their Roche lobes at advanced evolutionary stages, and the \COcore channel is entirely responsible for the DNS population. Due to minimal wind mass loss, the orbits remain compact and resilient to natal kick disruption, ensuring the bulk of these systems successfully merge within a Hubble time.
    
    \item $10^{-2}\Zsun \lesssim Z \lesssim \Zsun$: Both the \COcore and \Hecore channels lead to DNS formation. The highly evolved \COcore donors possess extended envelopes characterized by low binding energies. During the ensuing CE phase, these loosely bound envelopes are relatively easy to eject, demanding far less energy from the orbit. On the other hand, the younger \Hecore donors are less evolved and have tightly bound envelopes \citep[Fig. 10 in][]{Chattaraj+2026a}. Successful ejection requires greater orbital contraction, and if the systems survive, they produce tighter post-CE binaries. Both channels contribute to merging BNS, imprinting a double-peaked structure onto the delay time distribution, with the \Hecore channel peaking at shorter delay times \citep[see Fig. 6 in][]{Chattaraj+2026b}.
    
    \item $Z\gtrsim Z_\odot$: While both channels continue to form DNS, only the \Hecore channel produces merging systems. The extreme line-driven winds characteristic of the supergiant phase widens the pre-CE orbits of \COcore progenitors, pushing them out of the merging window. Consequently, only the \Hecore channel leads to merging BNS in this regime. In \citet{Chattaraj+2026a}, we demonstrated that this wind mass-loss-driven bifurcation at solar metallicity can reproduce the proposed subpopulations in the Milky Way DNS sample \citep{Andrews&Mandel2019}.    
\end{itemize}

Figure~\ref{fig:CO-HMS_RLO_populations} shows how the formation channels change with
metallicity, using slices of the \texttt{CO-HMS\_RLO} grid for a $1.43\,M_\odot$ NS and an H-rich donor at the onset of RLO. At all metallicities, most systems undergo unstable RLO leading to CE (diamonds), and stable RLO is rare (squares). The contours show the 75-th percentile distributions of where the \Hecore and \COcore subchannels of \texttt{MODEL03} lie in this parameter space. At $Z = 10^{-4}\,Z_\odot$, the \Hecore channel is entirely absent and all DNSs form through the \COcore channel. At $Z = 0.1\,Z_\odot$, both channels contribute; a secondary (model-dependent) pool of the \Hecore channel at shorter orbital periods accounts for $<20\%$ of the systems in that channel. At $Z = Z_\odot$, both channels are populated but only the \Hecore channel produces merging BNSs.

\section{Formation efficiency of binary neutron stars}
\label{sec:efficiency}

To construct the complete cosmic history of DNS formation, one must account for the star formation history of the Universe and its evolving metallicity. Consequently, determining whether DNS formation exhibits an intrinsic metallicity preference is essential.
Such a dependence not only determines their GW source population across cosmic time, but also the rates and host-galaxy demographics of their associated transients, which is essential for making predictions for current and next-generation GW detectors \citep{2010CQGra..27s4002P, Reitze+2019} and time-domain surveys such as Rubin-LSST \citep{2019ApJ...873..111I}.

\begin{figure*}
    \centering
    \includegraphics[width=0.85\linewidth]{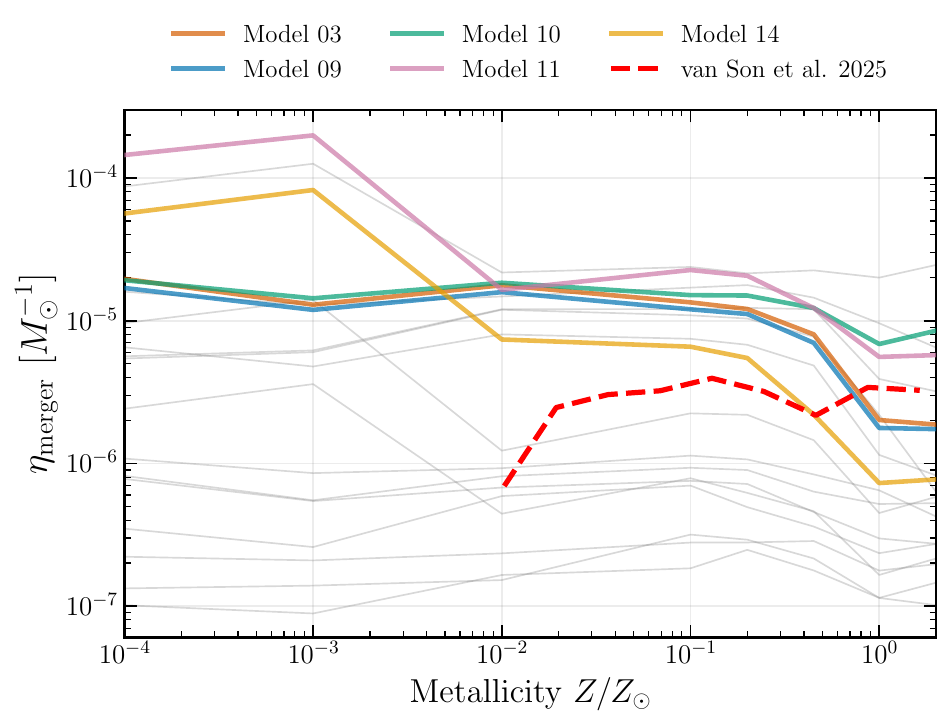}
    \caption{
    The intrinsic BNS merger efficiency per unit stellar mass formed as a function of progenitor metallicities for the full set of population synthesis models considered in this work (Table \ref{tab:model_populations}). Individual \texttt{POSYDON} model variations exploring uncertainties in binary physics are shown in thin grey lines, and our observationally calibrated subset is highlighted in color. For comparison, the dashed red line denotes the meta-analysis from \citet{vanSon+2025}, which consists of predictions from rapid population synthesis studies and only covers $Z \ge 10^{-2}\Zsun$. Our observationally calibrated models
    predict a decline in the efficiency toward higher metallicities, unlike the flatter trend of the meta-analysis over the metallicity range in common.
    }
    \label{fig:efficiency}
\end{figure*}

\begin{figure*}
    \centering
    \includegraphics[width=1.00\linewidth]{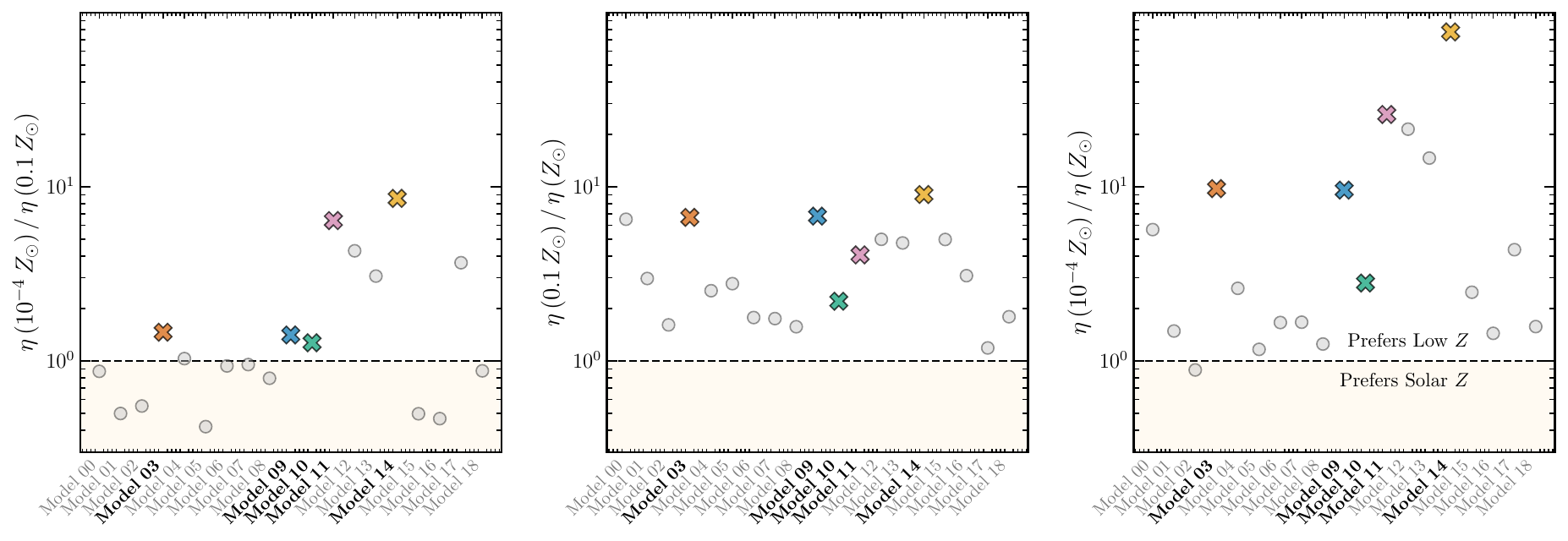}
    \caption{
    The relative BNS merger efficiency across different progenitor metallicity regimes for all model populations. The panels provide the ratio of merger efficiencies between $10^{-4}\,Z_\odot$ and $0.1\,Z_\odot$ (\textit{left}), $0.1\,Z_\odot$ and $Z_\odot$ (\textit{middle}), and across the entire metallicity range considered in this work from $10^{-4}\,Z_\odot$ to $Z_\odot$ (\textit{right}). The horizontal dashed line indicates a value of unity, where the formation efficiency is equal between the two compared metallicities; models falling above the line favors BNSs merging in the relatively lower metallicity environment. \textit{Left panel:} The scatter of all models above and below unity indicates that efficiencies in the very metal-poor regime are sensitive to binary physics uncertainties, though all observationally calibrated models favor $10^{-4}\,Z_\odot$. \textit{Middle panel:} In stark contrast, every model falls strictly above unity, demonstrating a universal preference for $0.1\,Z_\odot$ over $Z_\odot$. This is driven by the suppression of the \COcore channel in forming merging BNS toward high metallicity due to extreme supergiant wind mass loss and orbital widening, increasing the binary's susceptibility to natal kick disruption. \textit{Right panel:} Across the full metallicity range, almost all models favor metal-poor environments. Our observationally calibrated subset predicts an efficiency enhancement of roughly an order of magnitude at $10^{-4}\,Z_\odot$ compared to $Z_\odot$.
    }
    \label{fig:efficiency_ratios}
\end{figure*}

\subsection{The formation efficiency as a function of metallicity}
\label{subsec:eta}

BNS formation has previously been argued to be metallicity-agnostic \citep{Giacobbo&Mapelli2018, Klencki+2018, Neijssel+2019, Santoliquido+2021, vanSon+2025}, making them ideal candidates for ``spectral siren" cosmology \citep{Ezquiaga&Holz2022, SKRoy+2025}. However, recent analyses of the Milky Way $r$-process abundances \citep{Lee+2026} and the properties of sGRB host galaxies \citep{Nugent+2022} indicate that the BNS merger efficiency must evolve, increasing towards lower metallicity \citep[see also][]{Gallegos-Garcia+2023}. The detailed treatment of binary evolution available in \posydon provides an opportunity to revisit this question.

We define the intrinsic merger efficiency at a specific metallicity $\eta_{\rm merger}\,(Z)$, as the number of merging BNSs formed per unit star-forming mass $M$ at that metallicity,
\begin{equation}
    \eta_{\rm merger}\,(Z) = \frac{dN}{dM}. 
\end{equation}
Figure \ref{fig:efficiency} shows $\eta_{\rm merger}\,(Z)$ evaluated for all our model populations (grey lines) and the observationally calibrated subset (colored lines) across the entire metallicity range explored in this work. 
Our models show a large variation in the exact value of the efficiency at different metallicities, indicating how sensitive BNS formation is to the underlying binary physics assumptions. However, while the metallicity dependence across our full suite of models ranges from uniform formation efficiency to a strong preference for low metallicities, all models unambiguously predict a higher efficiency at $10^{-4}\Zsun$ (our lowest metallicity) and $0.1\Zsun$ compared to $\Zsun$ (middle and rights panels of Figure \ref{fig:efficiency_ratios}).
In particular, our observationally calibrated subset of models shows, on average, at least one order-of-magnitude enhancement in the efficiency going from $\Zsun$ down to $10^{-4}\Zsun$.
This low-metallicity preference stands in clear disagreement with the efficiency reported by \citet[][]{vanSon+2025}, overplotted as a dashed red line, whose meta-analysis compiles results from multiple rapid population synthesis studies. We note that the \citet{vanSon+2025} study does not extend below $Z = 10^{-2}\,Z_\odot$, whereas
our models reach $10^{-4}\,Z_\odot$. The comparison therefore cannot test the enhanced efficiencies that some of our calibrated models predict at the lowest metallicities (e.g., \texttt{MODELS 11} and \texttt{14}).

The broad evolutionary factors responsible for a low-metallicity preference of the efficiency are as follows. First, the weaker wind mass loss at low metallicities produces tighter binary orbits (see discussion in Section \ref{subsec:CO-HMS_RLO}) which is accompanied by an increased survivability from natal kick disruption. The tight orbits also imply short merger times on a population level, leading to more merging BNSs at lower metallicities \citep[in agreement with][]{Gallegos-Garcia+2023}.

Second, natal kicks predominantly disrupt wide binaries during the first SN \citep[e.g.,][]{Renzo+2019, Wagg+2025}, but play a comparatively minor role in the second SN. By this stage, multiple rounds of mass transfer episodes have significantly tightened the binary, and moreover, the second-born NS typically receives weaker natal kicks \citep[see][]{Disberg+2024, Chattaraj+2026a, Disberg+2026}. The adopted kick model therefore plays a role; models calibrated to isolated pulsar observations (Maxwellian: \citealt{Hobbs+2005} and lognormal: \citealt{Disberg&Mandel2025} in Table \ref{tab:model_populations}) produce stronger natal kicks that preferentially disrupt wide-orbit systems, compared to the ``physics-based" natal kick prescriptions. 

Third, stripped star winds govern the late-stage orbital evolution by removing mass at the specific angular momentum of the donor \citep{Ivanova+2003, Dewi&Pols2003}, thereby expanding the orbit and making the binary susceptible to natal kick disruption. This effect is most pronounced at metallicities $>0.1\Zsun$, which can be seen by the factor of $\eta(0.1\Zsun)/\eta(\Zsun) \sim 7$ for \texttt{MODEL03}, whereas only $\sim 2$ for \texttt{MODEL10}, which adopts the ``low He star wind" implementation. Reducing the strength of He star winds at high metallicity allows more systems to have tighter orbits and survive natal kick disruption.

\subsection{Dominant physics across metallicity regimes}
\label{subsec:eta_met_regimes}

\begin{figure*}
    \centering
    \begin{subfigure}[b]{0.48\linewidth}
        \centering
        \includegraphics[width=\linewidth]{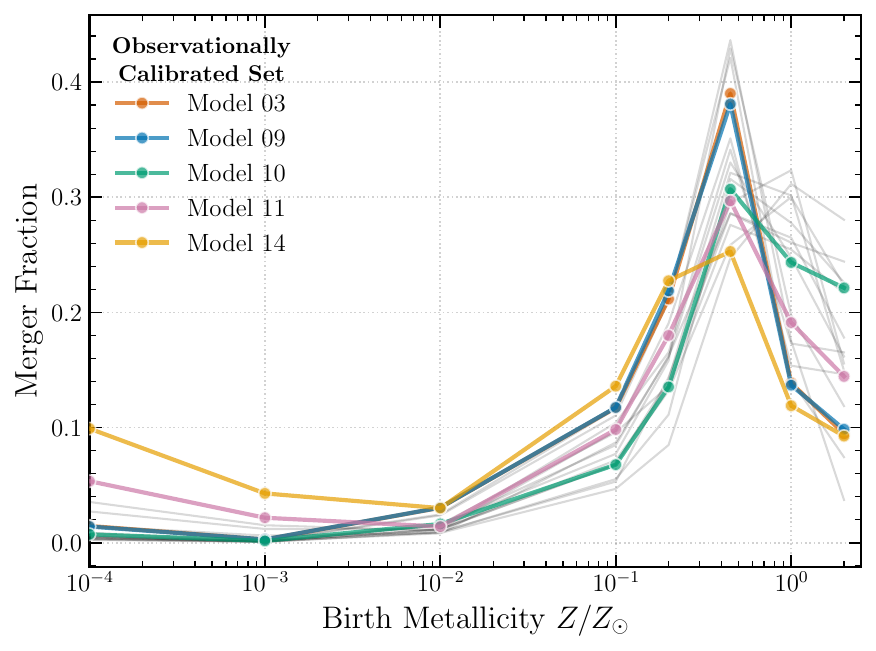}
        \caption{SFR: IllustrisTNG}
        \label{fig:frac_merger_illustrisTNG}
    \end{subfigure}
    \hfill
    \begin{subfigure}[b]{0.48\linewidth}
        \centering
        \includegraphics[width=\linewidth]{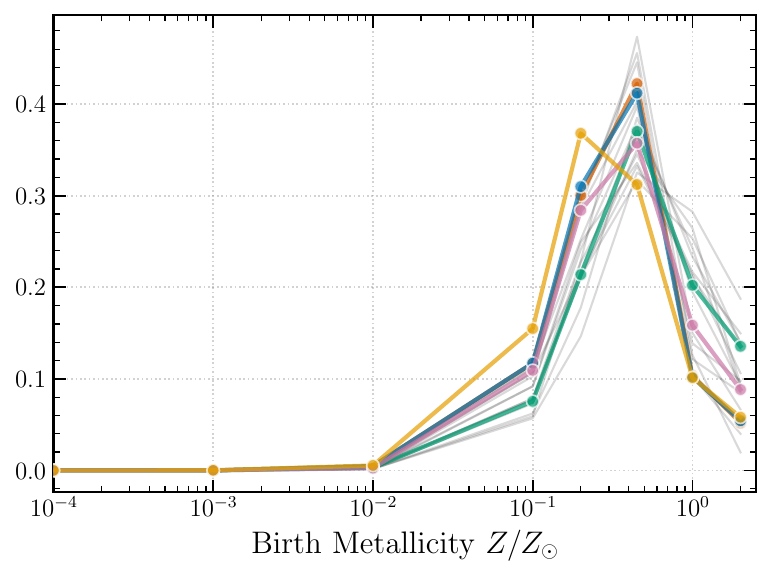}
        \caption{SFR: Madau+Fragos17}
        \label{fig:frac_merger_MF17}
    \end{subfigure}
    \caption{
    The birth metallicity distribution of cosmological BNS that merge in a Hubble time. The intrinsic formation efficiencies from our \texttt{POSYDON} models (Figure \ref{fig:efficiency}) are convolved with the cosmic star formation history from \texttt{IllustrisTNG} (left) and \texttt{MF17} (right). Despite the intrinsic preference for BNS mergers to form from low metallicity progenitors, both star formation histories demonstrate that the cosmological merging population is overwhelmingly dominated by progenitors born in the $0.1-1\,Z_\odot$ regime, with significant contribution from supersolar metallicities as well.
    The enhanced fraction of mergers below $10^{-2}\,Z_\odot$ in the left panel is a direct consequence of the \texttt{IllustrisTNG} model predicting a larger volume of low-metallicity star formation relative to analytical models like \citealt{Madau&Fragos2017} (see Figure \ref{fig:total_Msun_by_met}).
    }
    \label{fig:frac_merger}
\end{figure*}

The origin of the metallicity dependence becomes clearer when the progenitors are separated into two broad metallicity regimes, between $\Zsun - 0.1\Zsun$ and $0.1\Zsun - 10^{-4}\Zsun$ (see left and middle panels of Figure \ref{fig:efficiency_ratios}).
As metallicity decreases from $\Zsun$ down to roughly $0.1Z_\odot$, the models generally show a steep rise in the efficiency, which can be understood in two ways. First, at metallicities $\gtrsim 0.1\Zsun$, the \COcore formation channel is progressively shifted toward wider orbital periods, reducing the fraction of systems that form merging BNSs. This effect becomes increasingly pronounced with increasing metallicity, such that at $\Zsun$ the \COcore channel is pushed out of the merging window and only the \Hecore channel produces merging BNSs. This naturaly decreases the number of merging BNSs. Second, a combination of metallicity-dependent stellar winds and natal kicks suppresses the efficiency at high metallicity. At near-solar metallicities, strong supergiant winds from H-rich stars and line-driven winds from stripped helium stars widens the binary orbit, making them susceptible to disruption by natal kicks, resulting in a lower efficiency.

Below $\simeq 0.1\Zsun$, the stellar winds weaken, allowing tighter orbits that are more resistant to natal kick disruption. Therefore, natal kicks play a less impactful role. In this regime ($10^{-4}\,\Zsun \lesssim Z \lesssim 0.1\Zsun$), the BNS yield becomes sensitive to the physics governing the stellar collapse itself. For example, we find that the choice of SN prescription impacts the yield; the S16 model \citep{Sukhbold+2016} yields a higher fraction of successful explosions compared to the PS20 model \citep{Patton&Sukhbold2020}, directly increasing the merger efficiency. This is explicitly demonstrated by the increased relative efficiency $\eta\,(10^{-4}\,\Zsun)/\eta\,(\Zsun)$ in \texttt{MODEL03} compared to a decrease in \texttt{MODEL00}, models which are otherwise completely identical save for their SN prescriptions.

Furthermore, the weak winds at low metallicity allows stars to retain more of their initial mass, lowering the minimum ZAMS mass required to form a NS \citep[which leads to different formation timescales across metallicities, see Fig. 5 in][]{Chattaraj+2026b}. Since the IMF is bottom-heavy, the available number of progenitors increases at low metallicity. For example, assuming an IMF from \citet{Kroupa2001}, $\xi(m) \propto m^{-2.3}$, the total number of NS progenitors increases by a factor of $\sim 1.6-1.9$ if the minimum ZAMS mass required to form a NS decreases from $8\Msun$ at $\Zsun$ to $5.5\Msun$ at $10^{-4}\Zsun$. This naturally compunds to the enhancement in formation efficiency in metal-poor environments.

Additionally, uncertainties in modeling the CE phase are simplified in this regime ($\lesssim0.1\Zsun$). Standard CE modeling depends on three uncertain parameters: the ejection efficiency $\alpha_{\rm CE}$, the envelope binding energy parameter $\lambda$, and the core-envelope boundary defined by the hydrogen mass fraction $X_{\rm H}$. In \posydon, we bypass uncertainties from analytical fits for $\lambda$ by computing it directly from the detailed stellar structure of the donor star. Additionally, the suppressed HG radial expansion at low metallicities delays the CE phase until the donor is highly evolved (Section \ref{subsubsec:radius_evol}). By this stage, a steep density gradient has formed at the core-envelope boundary, and varying $X_{\rm H}$ produces negligible change in the mass coordinate of the core \citep[compare the top panels in Fig. 9 of][]{Chattaraj+2026a}. Thus, uncertainties in the CE modeling at low metallicities from a population synthesis standpoint is effectively reduced to a single parameter $\alpha_{\rm CE}$.

\subsection{Metallicity contribution to the cosmological population}
\label{subsec:cosmic_contribution}

While Figure \ref{fig:efficiency} demonstrates an increased BNS formation efficiency at lower metallicities, $\eta_{\rm merger}\,(Z)$ alone does not determine the metallicity distribution of observed BNS mergers. The overall contribution depends on: 1) the relative amount of star formation at different metallicities, and 2) the fraction of BNS systems that merge. To quantify these effects, we calculate the total comoving mass density $M_{\mathrm{form}}(Z)$ of stars formed at each metallicity, 
\begin{equation}
    M_{\mathrm{form}}(Z_{\rm j}) = \int_{0}^{z_{\mathrm{max}}} \psi(z) \, f_{\mathrm{SFR}}(Z_{\rm j}, z) \, \left|\frac{dt}{dz}\right| \, dz
    \label{eq:sf_per_met}
\end{equation}
where $\psi(z)$ is the cosmic star formation rate density and $f_{\mathrm{SFR}}(Z_{\rm j}, z)$ is the fraction of the total SFR at redshift $z$ that occurs in the metallicity bin centered on $Z_j$, whose edges lie at the logarithmic midpoints between adjacent grid metallicities.
We model $\psi(z)$ using the prescriptions outlined in Section \ref{subsec:setup}. Given that low-metallicity star formation in the Universe remains poorly constrained \citep{Chruslinska+2021}, we test four separate SFH models in \posydon: the cosmological \texttt{IllustrisTNG} simulation, and empirical models from \citet{Madau&Fragos2017}, \citet{Madau&Dickinson2014}, and \citet{Neijssel+2019}. As seen in Figure \ref{fig:total_Msun_by_met}, these models generally agree for $Z>10^{-2}~Z_{\odot}$, but diverge at lower metallicities.\footnote{For a detailed analysis on how analytical SFH models compare to \texttt{IllustrisTNG} simulations, see \citet{Levina+2026}.} Given the relatively close agreement of the three empirical models, we restrict our subsequent calculations to two representative models: the \texttt{IllustrisTNG} simulation and \texttt{MF17} model \citep{Madau&Fragos2017}, which together capture the relevant systematic uncertainties in cosmic star formation.

Next, we multiply $M_{\mathrm{form}}(Z)$ with $\eta_{\rm merger}\,(Z)$ to find the combined contribution of BNS mergers from different metallicities. 
We note that this product gives the birth metallicity distribution of all
merging BNSs produced over the cosmic SFH, not that of mergers at a particular redshift, which would require convolving with the delay time distribution (as we
do for the total merger rate in Section~\ref{sec:merger_rates}). With this caveat, since the delay times of our merging BNSs are predominantly short \citep[$\simeq \mathcal{O}(100\,$Myr);][]{Chattaraj+2026b}, our results closely approximates the metallicity distribution of all BNSs integrated throughout cosmic history that have merged by $z=0$.

Figure \ref{fig:frac_merger} shows results of this convolution for all model populations explored in this work. We find that the sheer volume of star formation across cosmic history is overwhelmingly concentrated in metal-rich environments \citep[$>0.1\Zsun$; see Fig. 6 in][]{vanSon+2023}. Therefore, despite the enhanced intrinsic formation efficiency at low metallicities, the fractional contribution to the cosmic merging BNS population drops precipitously for progenitors born below $Z \approx 0.1\,Z_\odot$ across all our model variations. For our observationally calibrated subset, progenitors born at $0.1$--$1\,Z_\odot$ account for $\approx 75$--$85\%$ of merging BNSs, with the single largest contribution ($\approx30$--$40\%$) coming from the $0.45\,Z_\odot$ bin. Progenitors
born at $Z \lesssim 10^{-2}\,Z_\odot$ contribute only a few percent, except for \texttt{MODELS 11} and \texttt{14} under \texttt{IllustrisTNG} ($\approx 10$-$15\%$), whose high efficiencies at the lowest metallicities (Figure~\ref{fig:efficiency}) combine with the larger amount of metal-poor star formation in that SFH model. Ultimately, while nature prefers low metallicity environments for BNS formation on a per-stellar-mass basis, the cosmic census of merging BNS is dominated by the prolific, albeit less efficient, near-solar metallicity star-formation of the Universe, specifically peaking between $0.1\,Z_\odot-Z_\odot$, with significant contributions from supersolar metallicities. 

If most short GRBs originate from BNS mergers, their host galaxies provide an observational test of our prediction. \citet{Nugent+2022} report that $\approx84\%$ of short GRBs occur in star-forming galaxies. With the BNS merger rate tracking cosmic SFH (see Section \ref{subsec:trends_merger_rates}), 
the present-day metallicity of a star-forming host may be a reasonable approximation for the birth metallicity of the progenitor stars. SED modeling of the larger sample\footnote{We note that host identification is
biased against faint, low-mass galaxies; and after correcting for completeness, $25-44\%$ of Swift short GRBs may originate at $z>1$ \citep{Fong+2022}. These biases will affect the low-metallicity regime, at which our models predict limited contribution, so current host galaxy samples cannot test that part of our prediction.} of \citet{Nugent+2022} gives a median stellar metallicity of $\log(Z_*/Z_\odot)=-0.38^{+0.44}_{-0.42}$ (68\% range), corresponding to roughly $0.16-1.15\,Z_\odot$. For long delay times, the host no longer reflects the progenitor's birth metallicity, and the metallicity of the host's old stellar population becomes the more relevant
proxy. For example, NGC 4993, the host galaxy of GW170817, assembled half of its stellar mass $\sim$11\,Gyr ago, and spectral modeling of its central region gives a median metallicity of $\rm{[Fe/H]} = 0.08^{+0.02}_{-0.03}$ \citep{Blanchard+2017}. We thus conclude that the short GRB host sample is consistent with our predicted range of $\sim0.1-1\,Z_\odot$, with NGC 4993 lying at the slightly supersolar end, where our models predict a significant contribution as well.

\section{Merger rates}
\label{sec:merger_rates}

\begin{figure*}
    \includegraphics[width=0.5\textwidth]{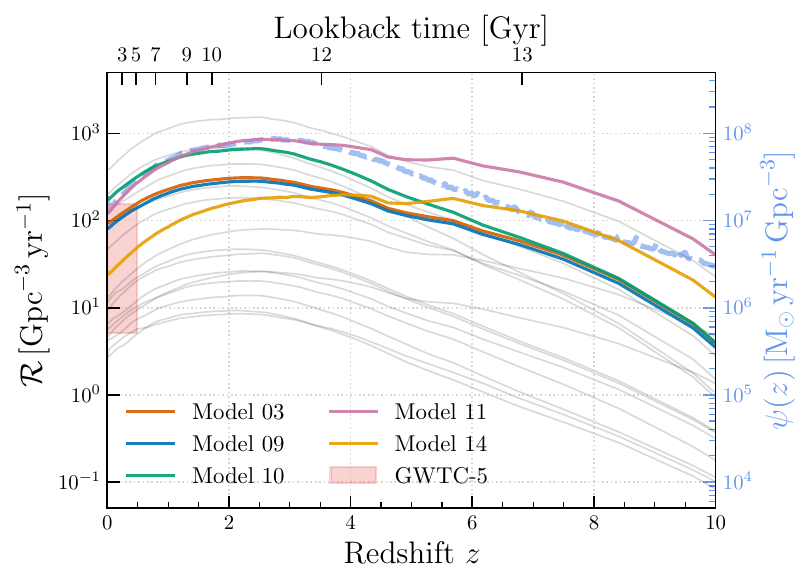}
    \includegraphics[width=0.5\textwidth]{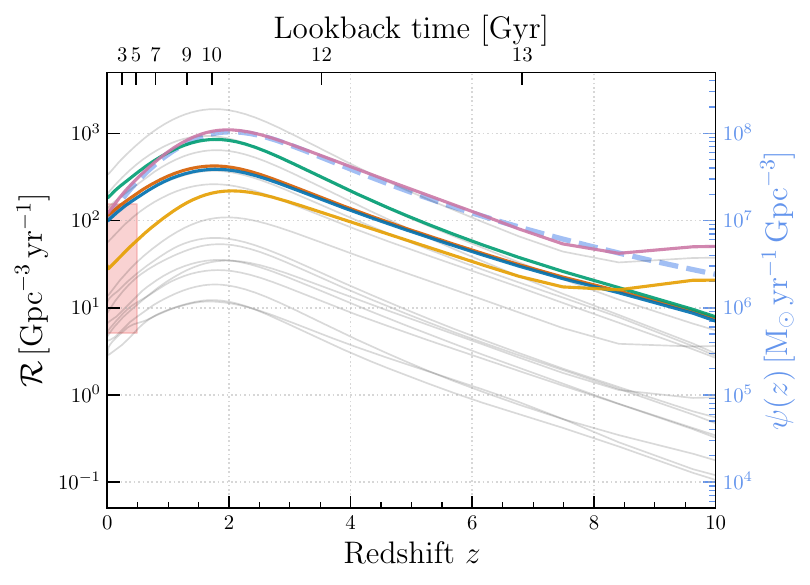}
     \caption{The BNS merger rate density evolution for all the population synthesis models in our work assuming two cosmic star formation histories: \texttt{IllustrisTNG} (left) and \texttt{MF17} (right). The grey lines representing our full set of population models with varying assumptions of binary physics exhibit a broad range in the rate evolution. Nevertheless, all models closely track the shape of the underlying SFH (blue dotted line), with merger rates typically peaking at $z\simeq2-3$ corresponding to cosmic noon. Additionally, our observationally calibrated subset (colored lines) broadly matches the latest local rate constraints from GWTC-5 (shaded in red, with an extended range up to $z=0.5$ for visualization purposes), although they lie near the upper end of the inferred range of $5.1-154.7\,\rm{Gpc^{-3}}\,\rm{yr^{-1}}$. This overlap may indicate that the observed Galactic DNS population need not be in tension with the merging BNS population inferred from GW observations.
     }
    \label{fig:merger_rates_all}
\end{figure*}

\begin{figure*}
    \centering
    \includegraphics[width=0.8\textwidth]{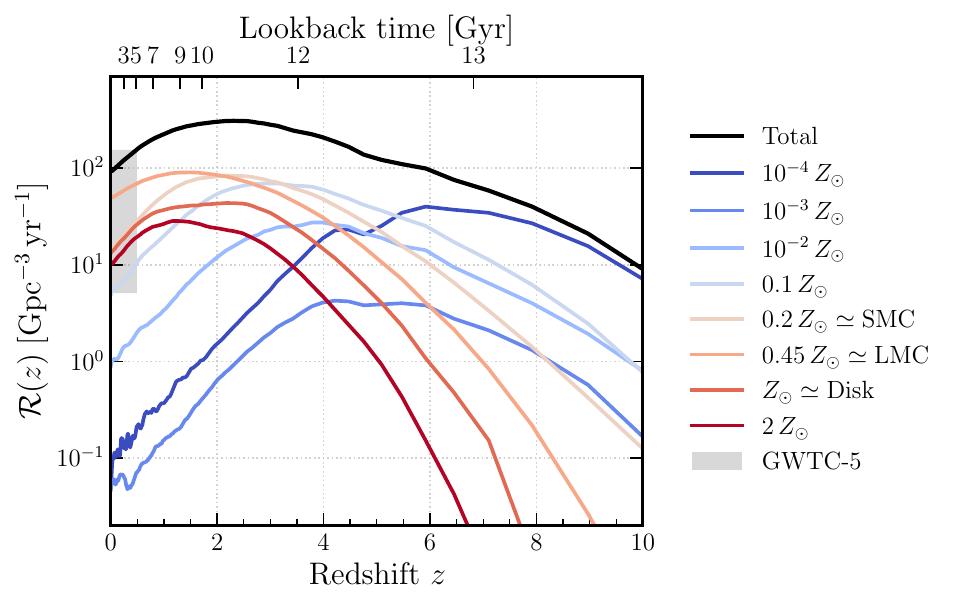}
    \caption{
    The redshift evolution of the BNS merger rate density for one of our observationally calibrated population $\texttt{MODEL03}$,  assuming the \texttt{IllustrisTNG} star formation history. The total rate (black) is broken down into contributions from eight birth metallicities and compared against the GWTC-5 constraints on the local rate (gray shaded region, which has been extended to $z=0.5$ for visualization). Three features of the cosmological BNS population can be inferred: (i) the local Universe ($z < 0.1$) is overwhelmingly dominated by progenitors born in the $0.1 - 2\,Z_\odot$ range; (ii) the contribution of metal-poor progenitors ($Z < 0.1\,Z_\odot$) peak at higher redshifts ($z\gtrsim3$); and finally, (iii) due to relatively short delay times, the the merger rate evolution broadly follows the cosmic star formation history, making BNS mergers potential tracers of cosmic chemical evolution.
    }
    \label{fig:merger_rate_by_met}
\end{figure*}

Recent studies have highlighted potential tensions between the BNS merger rates inferred from different observational probes \citep{Fishbach+2026, Kunnumkai+2026, Jin+2026}. To predict the cosmological evolution of the BNS merger rate density from the standpoint of isolated binary evolution, we convolve the delay times and merger efficiencies of our simulated binary populations with the SFHs adopted in this work (Section \ref{subsec:setup}).

\subsection{Trends in the cosmological merger rate evolution}
\label{subsec:trends_merger_rates}

Figure \ref{fig:merger_rates_all} shows the redshift evolution of the intrinsic merger rate density across our models, with the observationally calibrated subset highlighted by colored lines. We find that all models closely track the underlying star formation history (dashed curve, right axis), with a peak around cosmic noon at $z \approx 2$, and declining thereafter \citep[consistent with previous studies; see][]{Santoliquido+2021, Briel+2022, Iorio+2023, Boeksy+2024}. This trend implies that BNS delay times are predominantly short \citep[$\simeq \mathcal{O}(100\,$Myr);][]{Chattaraj+2026b}, allowing the merger rate to respond relatively quickly to changes in the star formation. The tested variations in binary physics assumptions represented by our models introduce a considerable spread in the predicted rates.
Additionally, the assumed cosmic SFH and mean metallicity evolution introduce small uncertainties in the predicted local ($z=0$) merger rates (see Table \ref{tab:rate_fits}). Thus, while binary physics remain the dominant source of uncertainty, accurate modeling of the cosmic SFH is required for precise BNS merger rate predictions.

To quantify the low redshift evolution of each model, we adopt the parameterization $\mathcal{R}(z) = \mathcal{R}_0\,(1+z)^{\kappa}$ commonly used in gravitational wave
population analyses \citep{Fishbach+2018}. We fix $\mathcal{R}_0$ to the simulated merger rate density at $z = 0$ and fit only $\kappa$, by linear least squares on $\ln(\mathcal{R}/\mathcal{R}_0)$ versus $\ln(1+z)$ over $0 \le z \le 1$. The resulting values are listed in Table~\ref{tab:rate_fits}. Across the full set of models, $\mathcal{R}_0$ spans more than two orders of magnitude
$\simeq 3$--$400$\,Gpc$^{-3}$\,yr$^{-1}$, while our predicted slopes range between $\kappa \simeq 1.2$--$2.4$. 
From Table \ref{tab:rate_fits}, we find that the slope $\kappa$ depends on the assumed SFH; every model evolves more steeply under \citet{Madau&Fragos2017} than \texttt{IllustrisTNG}.

The redshift evolution of the BNS merger rate is not yet
constrained by GW observations, with only two BNS merger detections and both at $z \lesssim 0.03$ \citep[GW170817 and GW190425;][]{LIGOGW170817+2017, LIGOGW190425+2020}. Our predicted slopes $\kappa$ can therefore be tested directly by next-generation detectors (Section~\ref{sec:detectability}). 
Furthermore, if BNS mergers are the dominant progenitors of sGRBs, $\mathcal{R}(z)$ predicted by our models sets the shape of their intrinsic rate density. The $(\mathcal{R}_0, \kappa)$ pairs in Table~\ref{tab:rate_fits} can therefore serve as theoretical inputs for sGRB luminosity function and redshift distribution studies \citep[e.g.,][]{Wanderman&Piran2015}, and for forecasts of joint GW--sGRB detections \citep[e.g.,][]{Ronchini+2022}. 

\begin{deluxetable}{lccccc}
\tablecaption{Power-law fits to the intrinsic BNS merger rate density, $\mathcal{R}(z) = \mathcal{R}_0\,(1+z)^{\kappa}$, over $0 \le z \le 1$, for the
\texttt{IllustrisTNG} and \texttt{MF17} SFH prescriptions.
\label{tab:rate_fits}}
\tablehead{
\colhead{} & \multicolumn{2}{c}{IllustrisTNG} & \colhead{} & \multicolumn{2}{c}{Madau\&Fragos} \\
\cline{2-3} \cline{5-6}
\colhead{Model} &
\colhead{$\mathcal{R}_0$} & \colhead{$\kappa$} & \colhead{} &
\colhead{$\mathcal{R}_0$} & \colhead{$\kappa$}
}
\startdata
\texttt{MODEL00}  & 47.1  & 1.59 & & 56.3  & 1.86 \\
\texttt{MODEL01}  & 5.4   & 1.57 & & 4.9   & 2.09 \\
\texttt{MODEL02}  & 3.0   & 1.30 & & 3.2   & 1.41 \\
\pmb{\texttt{MODEL03}} & 95.0  & 1.26 & & 117.7 & 1.42 \\
\texttt{MODEL04} & 6.5   & 1.54 & & 6.7 & 1.81 \\
\texttt{MODEL05} & 2.3   & 1.97 & & 2.4 & 2.35 \\
\texttt{MODEL06} & 195.0 & 1.63 & & 201.0 & 1.96 \\
\texttt{MODEL07} & 11.0  & 1.72 & & 11.5 & 1.94 \\
\texttt{MODEL08} & 4.2   & 1.34 & & 4.3 & 1.5 \\
\pmb{\texttt{MODEL09}} & 81.0  & 1.37 & & 99.2 & 1.56 \\
\pmb{\texttt{MODEL10}} & 182.2 & 1.39 & & 189.5 & 1.60 \\
\pmb{\texttt{MODEL11}} & 121.7 & 1.97 & & 123.5 & 2.28 \\
\texttt{MODEL12} & 13.7 & 1.78 & & 14.1 & 2.16 \\
\texttt{MODEL13} & 4.5 & 1.83 & & 4.7 & 2.13 \\
\pmb{\texttt{MODEL14}} & 25.5 & 1.65 & & 29.7 & 1.81 \\
\texttt{MODEL15} & 84.6  & 1.24 & & 102.3 & 1.52 \\
\texttt{MODEL16} & 125.9 & 1.33 & & 138.9 & 1.63 \\
\texttt{MODEL17} & 402.1 & 1.51 & & 352.2 & 1.90 \\
\texttt{MODEL18} & 9.3 & 1.88 & & 8.7 & 2.26 \\
\enddata
\tablecomments{$\mathcal{R}_0$ (in Gpc$^{-3}$\,yr$^{-1}$) is fixed to the simulated merger rate density at $z = 0$; only $\kappa$ is fitted.
Models in bold are the observationally calibrated subset.}
\end{deluxetable}

\subsection{Merger rates of the Galactic DNS-calibrated subset}
\label{subsec:galactic_dns_merger_rate}

As shown in Table \ref{tab:rate_fits}, the subset of models that best reproduces the Galactic DNSs (the observationally calibrated subset) predicts local merger rates that fall broadly within the observed credible interval of $5.1-154.7\,\rm{Gpc^{-3}}\,\rm{yr^{-1}}$ inferred from GWTC-5 \citep[the quoted range is the union of the 90\% credible intervals from the strongly-parameterized \fullpop and weakly-parameterized \pixelpop models;][]{gwtc5populations2026}, with the exception of \texttt{MODEL10}. Since our criteria for selecting the observationally calibrated subset are based entirely on binary evolution constraints (see Section \ref{subsec:galactic_dns}) and were not previously fitted to any merger rate predictions, this agreement represents an independent population synthesis prediction stemming from stellar physics.

However, a few caveats apply. First, the observed Galactic DNS sample is subject to the selection effects of radio pulsar surveys, including the finite radio lifetimes of pulsars and beaming corrections (see Section~5 of \citealt{Fishbach+2026}). A model that reproduces the Milky Way observations therefore need not be representative of the underlying population in every respect. Second, the detection volume for BNS mergers with current GW detectors is capable of probing only the very local Universe \citep{Abbott+2018}; and GWTC-5 constraints are derived from only two BNS detections, GW170817 \citep{LIGOGW170817+2017} and GW190425 \citep{LIGOGW190425+2020}. With these caveats in mind, our results may be interpreted as evidence that a single model of isolated binary evolution (with specific constraints to the underlying stellar physics) can simultaneously reproduce the observed properties of Milky Way DNSs while predicting a local merger rate consistent with GWTC-5 constraints. We thus hypothesize that the local DNS population seen in radio pulsar surveys need not be in fundamental tension with the cosmological GW-observed population.

\subsection{Metallicity contribution to the redshift evolution of the merger rate}

We further decompose the redshift evolution of the total merger rate density $\mathcal{R}(z)$ by the birth metallicity of the progenitor for our observationally calibrated \texttt{MODEL03} in Figure \ref{fig:merger_rate_by_met}.
In the local Universe ($z < 0.1$), the BNS merger rate is overwhelmingly dominated by progenitor systems born at $0.1\,\Zsun-2\,\Zsun$. These systems must therefore form the vast majority of events detectable by current-generation GW observatories. Additionally, the dominance of $Z \gtrsim 0.1\,Z_\odot$ progenitors persists out to $z \approx 2$, covering the range of most sGRBs with measured redshifts \citep{Nugent+2022}. In contrast, the contribution from metal-poor progenitors ($Z < 0.1\,\Zsun$) is negligible at low redshifts; instead, their merger rates peak much earlier in the Universe, at redshifts $z \gtrsim 3$.

This behavior follows from the combination of the Universe's enrichment history and the relatively short BNS delay times \citep{Chattaraj+2026b}. Since the mergers occur relatively soon after their progenitor systems form, the metallicity distribution of $\mathcal{R}(z)$ retains a strong imprint of the metallicity distribution of star formation at the corresponding cosmic epoch. In other words, the BNS population should trace star formation from $Z \gtrsim0.1\Zsun$ out to $z \approx 2$, whereas high-redshift mergers will likely probe the metal-poor stellar populations characteristic of the early Universe, which remains relatively unconstrained.

\section{Mass Distribution}
\label{sec:masses_and_detectability}

The intrinsic BNS mass distribution is a critical observational constraint. Yet, the core-collapse prescriptions currently used in population synthesis frameworks to map pre-SN cores to remnant masses is fraught with uncertainties (e.g., the monotonicity of this relationship remains an active area of research; see \citealt{Mandel&Muller2020, Burrows+2024, Karim+2026}). Moreover, these prescriptions may be insufficiently constrained to make definitive predictions across diverse metallicities. In this section, we first examine the robustness of the predicted BNS mass spectra across metallicity and redshift (Section \ref{subsec:mass_spectra}), and then bypass uncertainties in remnant mass prescriptions by looking at the pre-SN core masses instead (Section \ref{subsec:asymmetric_q}).

\subsection{Uncertainties and trends in the mass spectra}
\label{subsec:mass_spectra}

\begin{figure*}
    \centering
    \includegraphics[width=1.0\linewidth]{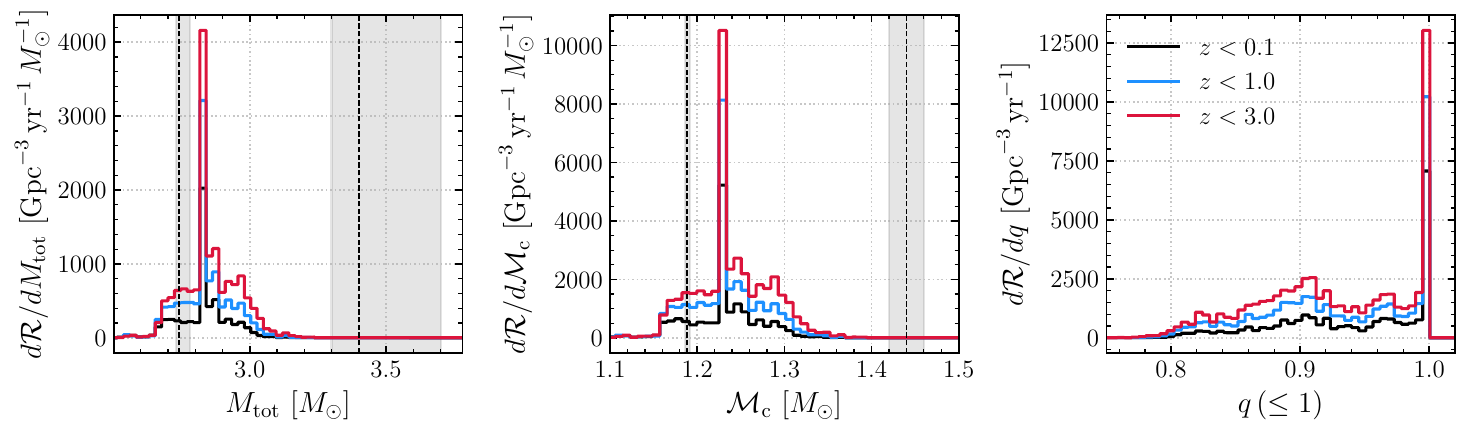}
    \includegraphics[width=1.0\linewidth]{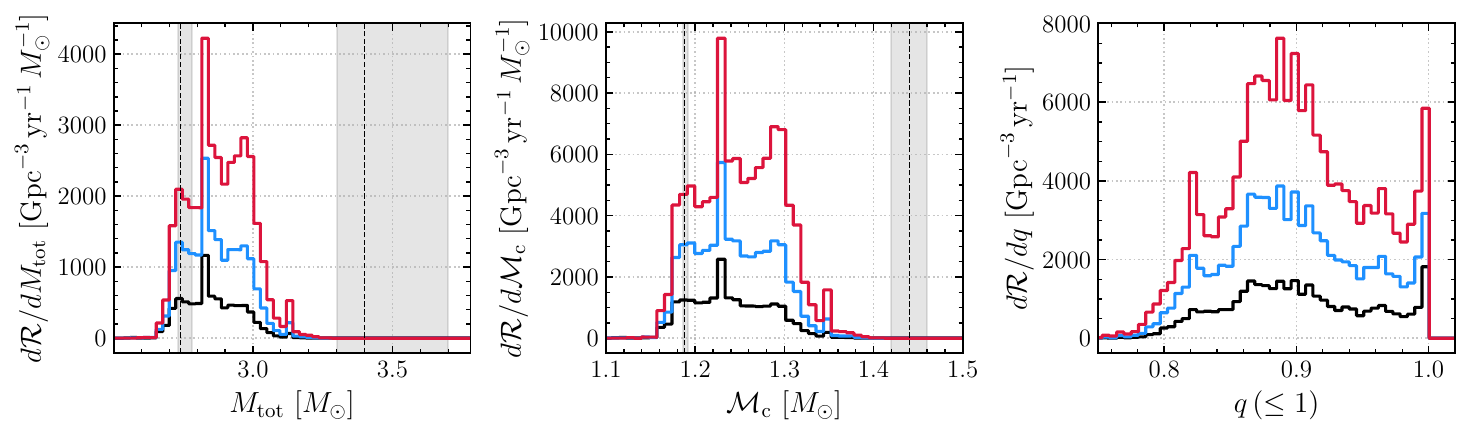}
    \caption{
     Distributions of the total mass ($M_{\rm tot}$, left), chirp mass ($\mathcal{M}_{\rm c}$, center), and mass ratio ($q$, right) in \texttt{MODEL03} (top row) and \texttt{MODEL11} (bottom row), respectively. Both models use the same SN prescription \citep{Sukhbold+2016} but differ in their CE treatment (Table~\ref{tab:model_populations}). We show the cumulative populations merging within redshift intervals $z < 0.1$ (black), $z < 1$ (blue), and $z < 3$ (red), where each interval includes the previous ones. The vertical dashed lines and gray shaded regions represent the median and 90\% credible intervals for the events GW170817 (lower values) and GW190425 (higher values).
     }
    \label{fig:mtot_mc_q}
\end{figure*}

Figure~\ref{fig:mtot_mc_q} shows the distributions of the total mass ($M_{\rm tot}$), chirp mass ($\mathcal{M}_c$), and mass ratio $q \ (\leq 1)$ of BNSs merging at $z<0.1, 1.0, 3.0$, for
\texttt{MODEL03} and \texttt{MODEL11}.
Both models peak near $M_{\rm tot} \approx 2.8\,M_\odot$ and $\mathcal{M}_c \approx 1.23\,M_\odot$, consistent with GW170817, but produce essentially no
systems with $M_{\rm tot} \gtrsim 3.2\,M_\odot$, and therefore cannot account for GW190425. Notably, we find that although the contribution from each metallicity varies, the shapes of the resulting mass spectra do not exhibit strong metallicity dependence \citep[see also][]{Nair&Stevenson2026}. 
We note that all SN prescriptions adopted in this work are calibrated to progenitors at $\Zsun$, implicitly imposing no metallicity dependence. However, pre-SN structures are expected to vary with metallicity \citep{Limongi&Chieffi2018}. Such an implementation is beyond the scope of this current work, and we refer the reader to \citet{Maltsev+2025} for recent modeling efforts incorporating metallicity-dependent explodability.

When evaluating the impact of the core-collapse prescriptions, we find that the model from \citet{Fryer+2012} fails to accurately capture expected BNS masses \citep[in agreement with][]{Mapelli&Giacobbo2018}. Conversely, the S16 prescription \citep{Sukhbold+2016} performs significantly better. 
However, fixing the supernova engine does not eliminate uncertainties in the mass spectra. For example, \texttt{MODEL03} and \texttt{MODEL11} shown in Figure \ref{fig:mtot_mc_q} both utilize the \citet{Sukhbold+2016} model, but different CE prescriptions. While the models predict qualitatively similar allowed ranges for these parameters, their density distributions differ. For instance, \texttt{MODEL03} produces a highly peaked mass ratio distribution dominated by near-equal mergers ($q \approx 1$, right panel in top row), whereas \texttt{MODEL11} yields a much broader distribution with a primary peak centered around $q \approx 0.88-0.90$. Similarly, the chirp mass distribution in \texttt{MODEL11} shows a prominent secondary peak near $\mathcal{M}_{\mathrm{c}} \approx 1.3\,M_\odot$ and a broader total mass distribution extending beyond $3.0\,M_\odot$, features that are largely suppressed in \texttt{MODEL03}. These differences demonstrate that in addition to the chosen SN prescription, BNS mass predictions from population synthesis are sensitive to the underlying treatment of binary physics.

While the NS remnant mass derived from precollapse structure and abundance is believed to be metallicity-dependent \citep{Aguilera-Dena+2023, Boccioli&Fragione2024, Ugolini+2025}, we are currently unable to resolve any significant metallicity dependence to the BNS mass spectrum \citep[in agreement with][]{Mapelli+2019, Nair&Stevenson2026}. Given these systematic uncertainties, we caution against strictly interpreting or constraining theoretical BNS mass models against current observational catalogs.

\subsection{Metallicity-driven asymmetries in the pre-supernova progenitors}
\label{subsec:asymmetric_q}

\begin{figure}
    \centering
    \includegraphics[width=1.00\linewidth]{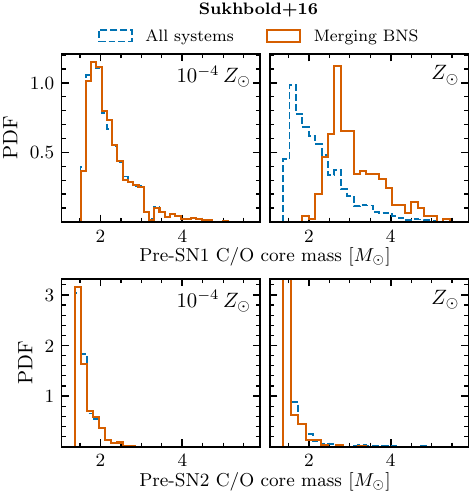}

    \vspace{4mm}
    \includegraphics[width=1.00\linewidth]{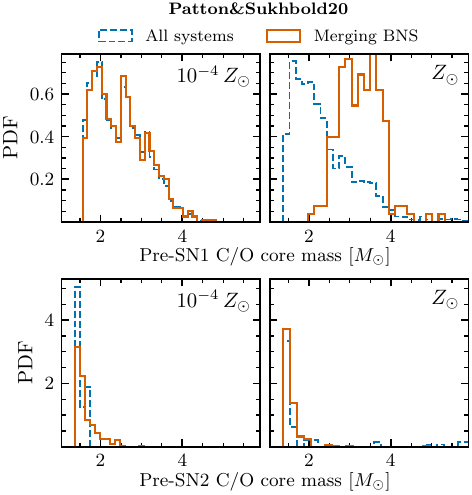}
    \caption{Normalized probability densities of the pre-SN C/O core masses of the first-born (SN1) and second-born (SN2) NSs, for \texttt{MODEL03} (top two rows, using \citealt{Sukhbold+2016}) and \texttt{MODEL00} (bottom two rows, using \citealt{Patton&Sukhbold2020}), at $10^{-4}\,Z_\odot$ (left) and $\Zsun$ (right). The dashed histograms show all DNS systems formed, while solid histograms denote the merging subset. For both prescriptions, the merging systems at $10^{-4}\,Z_\odot$ are representative of the full population. At $Z_\odot$, only systems with more massive C/O cores ($\gtrsim 2.5\,M_\odot$) evolve into merging BNSs, revealing a metallicity-dependent selection effect. The pre-SN2 C/O core masses fall between $\sim 1.4$--$2\,M_\odot$ in all cases.
    }
    \label{fig:pre-SN_core_mass}
\end{figure}

\begin{figure}
    \centering
    \includegraphics[width=1.00\linewidth]{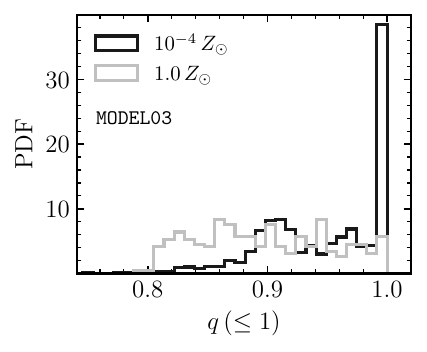}
    \caption{
    Normalized mass ratio ($q$) distributions of merging BNSs in \texttt{MODEL03} at $10^{-4}\,Z_\odot$ (black) and $Z_\odot$ (gray). The $10^{-4}\,Z_\odot$ population produces more symmetric BNS mergers, with $\gtrsim 80\%$ having $q \ge 0.9$, compared to $\simeq 50\%$ at $\Zsun$, which also lacks the pronounced peak near equal masses at ($q \simeq 1$).
    While assessing precise NS mass values stretches the limits of standard analytical SN prescriptions, this systematic shift toward asymmetry at high metallicities is a robust, metallicity-driven trend arising from binary evolution.
    }
    \label{fig:asymmteric_q}
\end{figure}




To circumvent uncertainties associated with the core-collapse process and remnant mass prescriptions, we now analyze the pre-SN core masses, assuming they are indicative of the birth NS mass distribution \citep[e.g.,][]{Laplace+2021}. Our motivation behind this approach is to investigate metallicity-dependent signatures in BNS masses that originate strictly from massive binary evolution.
Figure \ref{fig:pre-SN_core_mass} shows the distribution of the C/O core masses of the progenitors to the first-born (SN1) and second-born (SN2) NSs at $Z = 10^{-4}\,Z_\odot$ and $Z_\odot$ (although we show the C/O core mass, the effects discussed below apply to the He core mass as well). Furthermore, to establish trends that are agnostic to the SN model\footnote{Although we show pre-SN masses, they may not be independent of the SN prescription which sets the outcome of SN1 (NS or BH) in addition to the remnant mass, therefore determining which systems are classified as BNS.}, we show the distributions for \texttt{MODEL03} and \texttt{MODEL00}, which share identical binary physics assumptions except for their use of the  \citet{Sukhbold+2016} and \citet{Patton&Sukhbold2020} SN prescriptions, respectively.

At $10^{-4}\,Z_\odot$, the pre-SN1 C/O core masses for merging systems are broadly representative of the underlying total population (i.e., including the non-merging systems). However, at solar metallicity $Z_\odot$, a selection effect emerges: the non-merging systems are dominated by lower pre-SN1 C/O core masses, whereas the merging subset preferentially come from systems with more massive cores. This difference is a direct consequence of the strong stellar winds at high metallicities which simultaneously widens the orbit and strips the envelope, stalling core growth. Taken together, they lead to a population-level effect where systems with lower pre-SN1 core masses survive in wider orbits and are thus less likely to merge, leaving higher mass cores preferentially represented among the merging systems.This trend is prevalent across both SN prescriptions shown in Figure \ref{fig:pre-SN_core_mass}.

In contrast, the pre-SN2 core masses for the merging systems show little to no dependence on the birth metallicity (Figure \ref{fig:pre-SN_core_mass}). 
This is because by the time the secondary star collapses, it has typically been stripped of both its H-rich and He-rich envelopes through interaction with the first-born NS. These ultra-stripped cores explode as electron-capture supernovae or low-mass Fe core-collapse \citep{Tauris+2015} which naturally restricts the pre-SN2 cores to a very narrow, low-mass range ($\sim 1.4$--$2\,M_\odot$). This picture is qualtitavely consistent with the Galactic DNS population, which formed at near-solar
metallicity; the second-born NSs have systematically low masses and small inferred kicks, as expected from ultra-stripped progenitors; whereas the recycled, first-born NSs span a broader and higher range of masses \citep{Farrow+2019, Chattaraj+2026a}.

Taking together the effects of binary evolution on pre-SN1 and pre-SN2 core masses, we identify a metallicity-driven prediction for the intrinsic BNS mass spectrum. Since the second-born NS mass is confined to a narrow range of masses, the mass ratio ($q$) of the resulting binary is determined by the mass of the primary NS. Since merging BNSs at higher metallicities show more massive pre-SN1 cores, we expect these BNS mergers to intrinsically exhibit more asymmetric mass ratios than those formed at low metallicities. Indeed, the intrinsic mass ratio distributions for \texttt{MODEL03} shown in Figure \ref{fig:asymmteric_q} bear out this trend.\footnote{Unlike the pre-SN core masses, the mass ratios in Figure~\ref{fig:asymmteric_q} are
computed from NS masses, which require a remnant-mass prescription (here
\citealt{Sukhbold+2016}). Although it supports the trend we discuss, the exact BNS mass ratios are still model dependent and should be interpreted qualitatively.} 
While $\gtrsim 80\%$ of BNS mergers have $q \geq 0.9$ at $10^{-4}\Zsun$, this fraction decreases to $\simeq 50\%$ for the $\Zsun$ population \citep[see also][]{Andrews2020}, which lacks the distinct peak at near-equal mass mergers present at $10^{-4}\Zsun$. The consequences of this metallicity-dependent asymmetry in the BNS masses for the merger ejecta, and the resulting GRB and kilonova \citep{Gottlieb+2023, Gottlieb+2025} will be explored in future work.

\section{Detectability with next-generation gravitational wave observatories}
\label{sec:detectability}

\begin{figure*}
    \includegraphics[width=0.48\linewidth]{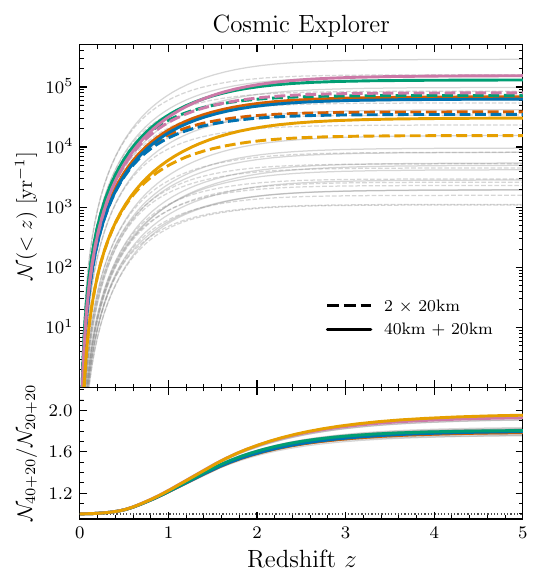}
    \includegraphics[width=0.48\linewidth]{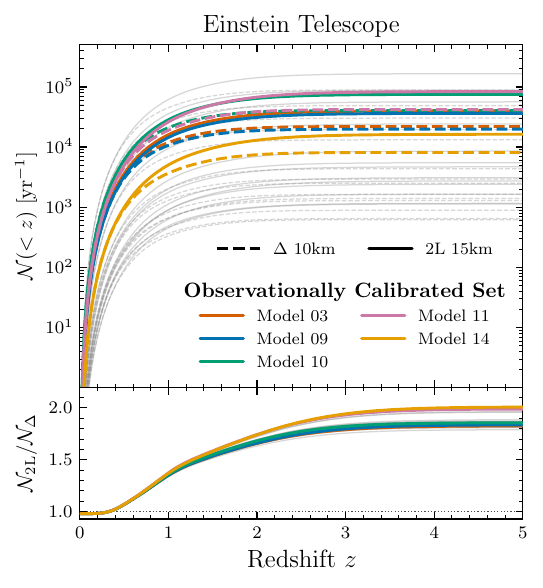}
    \caption{The cumulative number of detectable BNS mergers per observer year as a function of redshift for proposed 3G detectors, Cosmic Explorer (left) and the Einstein Telescope (right), is shown for all the population models considered in this work. 
    The observationally calibrated models are highlighted in color and the remaining are shown in gray.
    (\textit{Top panels}) For Cosmic Explorer, solid and dashed lines show the 40\,km\,+\,20\,km and 20\,km\,+\,20\,km networks; for the Einstein Telescope, they show the 2L configuration with 15\,km arms and the triangle with 10\,km arms. (\textit{Bottom panels}) Ratio of the cumulative detections of the more sensitive configuration to that of the less sensitive one. Our observationally calibrated models predict $\simeq 10^4$--$10^5$ detectable BNS mergers per year out to a redshift $z\sim2$ (see Table \ref{tab:model_populations}).}
    \label{fig:detectability}
\end{figure*}

The next generation (3G) of gravitational wave observatories like Cosmic Explorer and Einstein Telescope will possess the sensitivity required to detect BNS mergers out to and beyond redshifts probing the peak of cosmic star formation \citep{Sathyaprakash+2019, Reitze+2019, Maggiore+2020, Evans+2021, Kalogera+2021}. To evaluate the observable yield of our model populations within 3G detectors, we compute the BNS detection probability $P_{\rm det}(z)$ with \texttt{gwfast} \citep{gwfast2022a, gwfast2022b}, modeling each binary with source-frame masses $m_1 = m_2 = 1.4\,M_\odot$ using the \texttt{TaylorF2\_RestrictedPN}. 
At each redshift, we average over $10^4$ Monte Carlo draws of isotropic sky positions, orbital inclinations, polarizations, and coalescence times spanning a sidereal day, including the Earth's rotation during the hours-long signal. Each interferometer is assumed to operate independently 85\% of the time \citep{Branchesi+2023}, and a source is detected if its network SNR is $\rho_{\rm net} \geq 8$. We use the intrinsic merger rate density $\mathcal{R}(z')$ for each population model derived from the \citet{Madau&Fragos2017} SFH to calculate the cumulative detection rates per observer year,
\begin{equation}
    \mathcal{N}(<z) = \int_0^z \frac{R\,(z')}{1+z'} \frac{d V_c}{d z'} \, P_{\mathrm{det}}(z') \, d z'
\end{equation}
where  $d V_c / d z$ is the differential comoving volume and the factor $(1+z)^{-1}$ accounts for cosmic time dilation between the source and observer frames assuming a standard flat $\Lambda$CDM cosmology \citep{Planck+2020}.
For Einstein Telescope, we compare a 10\,km triangular configuration with two 15\,km L-shaped detectors misaligned by $45^\circ$. For Cosmic Explorer, we compare two 20\,km detectors with a 40\,km\,+\,20\,km network. Figure \ref{fig:detectability} shows the results for all our population models, with the observational calibrated ones highlighted in color.

Across the full set of population models, both Cosmic Explorer and Einstein Telescope networks predict detection rates varying over two orders of magnitude, between $\simeq 10^3 - 10^5$ per observer year \citep[see][]{Baibhav+2019, Ronchini+2022, Steinle+2025, Koehn+2026}. This spread reflects the sensitivity of the observable cosmological BNS population to the compounded uncertainties from intrinsic metallicity-dependent binary physics and the cosmic SFH used to predict their merger rate densities \citep[see also][]{Wagg2022, Toubiana&Dvorkin2026}. 

For Cosmic Explorer, we find that the $40\,\mathrm{km} + 20\,\mathrm{km}$ detector network outperforms a $20\,\mathrm{km} + 20\,\mathrm{km}$ network by a factor of $\simeq 2$ in the cumulative number of detectable BNS mergers \citep{Evans+2021, Evans+2023}.
Similarly, for Einstein Telescope, the 2L shaped configuration with 15\,km arms detects roughly twice as many BNS mergers as the triangular shape with 10km arms \citep[in close agreement with][see their Table 42]{Branchesi+2023}. For both observatories, the competing configurations yield nearly identical cumulative rates for nearby mergers ($z < 0.5$), implying that they are detected with high efficiency by every design considered (bottom panels of Figure \ref{fig:detectability}). The gains of the more sensitive designs ($40\,\mathrm{km} + 20\,\mathrm{km}$ for Cosmic Explorer, and 2L 15km for Einstein Telescope) grow at $z \gtrsim 1$ and levels off beyond $z \simeq 3$, implying that their advantage lies in sampling BNS mergers near and beyond the peak of cosmic star formation, the regime most critical for constraining the the delay time distribution and the metallicity dependence of BNS formation.

\section{Conclusions}
\label{sec:conclusions}

We explored the formation of DNSs across a cosmological range of metallicities using the detailed binary evolution code \posydon. By modeling the metallicity-modulated evolution of single- and binary-stars, along with full-scale population synthesis, we connected intrinsic massive stellar evolution to the properties of the broader cosmological DNS population. Building upon our previous study at solar-metallicity \citep{Chattaraj+2026a}, we identified an ``observationally calibrated'' subset of models that provides the best representation of the observed Galactic DNS sample. We use this subset to interpret how binary evolution models consistent with the local Galactic population would evolve across the diverse metallicity environments throughout cosmic history. Our main conclusions are summarized below:

\begin{enumerate}
    \item The CE channel is the dominant pathway to form DNSs across all metallicities. Its relative contribution increases towards low metallicity; such that all DNS formation in our models at $10^{-4}\Zsun$ proceeds through the CE channel (Section \ref{subsec:all_form_channels}).
    
    \item Metallicity determines the available CE subchannels through its impact on the radial evolution of the donor. At $Z \lesssim 10^{-2}\Zsun$, suppressed radial expansion during the HG phase delays RLO until later evolutionary stages, when the donor has developed a carbon-oxygen core, restricting DNS formation to the \COcore channel. Conversely, at $Z \gtrsim 0.1\Zsun$, donors undergo sufficient expansion during the HG phase to fill their Roche lobes, allowing both the \Hecore and \COcore formation channels (Section \ref{subsubsec:radius_evol}).    
    
    \item Metallicity-dependent supergiant winds further influences the pre-CE orbital evolution. At $Z \lesssim10^{-2}\Zsun$, the donors evolve into compact, blue supergiants, with wind mass-loss rates that are roughly two orders-of-magnitude weaker than those of their higher metallicity counterparts ($\gtrsim 0.1\Zsun$), which evolve into cooler, red supergiants (Section \ref{subsubsec:wind_mass_loss}). This leads to substantially different pre-CE orbital expansion, producing a distinct outcropping of models in the parameter space of our H-rich star$-$NS binary grids at $Z \gtrsim 0.1\Zsun$ (Figure \ref{fig:CO-HMS_RLO_multi_panel}). For $Z\gtrsim \Zsun$, these winds become strong enough to widen the \COcore channel such that it no longer produces merging BNSs (Sections \ref{subsec:single_star} and \ref{subsec:CO-HMS_RLO}).
    
    \item The intrinsic BNS merger efficiency increases toward low metallicities across varying prescriptions for binary evolution (Section \ref{subsec:eta}). In particular, our observationally calibrated subset predicts roughly one order-of-magnitude enhancement in the merger efficiency at $10^{-4} Z_{\odot}$ compared to $\Zsun$ (see Figure \ref{fig:efficiency}). 
    
    \item The physical processes determining the BNS merger efficiency varies across different metallicity regimes (Section \ref{subsec:eta_met_regimes}).
    
    \begin{itemize}
        \item At $Z\gtrsim0.1\Zsun$, a combination of wind-driven orbital widening and natal kick disruption, along with the progressive suppression of merging BNSs from the \COcore channel, plays a dominant role in reducing the BNS merger efficiency. Consequently, all of our models predict a higher efficiency at $0.1\Zsun$ than at $\Zsun$ (Figure \ref{fig:efficiency_ratios}).

        \item Below $\simeq 0.1\Zsun$, the stellar winds weaken substantially and BNS efficiency becomes more sensitive to the core-collapse prescription and the CE energy efficiency parameter $\alpha_{\rm CE}$. Although our full set of population models do not unanimously agree whether the efficiency continues to increase from $0.1\Zsun$ to $10^{-4}\Zsun$, the observationally calibrated subset still retains a preference for the lower metallicity (Figure \ref{fig:efficiency_ratios}).
    \end{itemize}
    
    \item Despite the enhanced intrinsic efficiency at low metallicity, our populations convolved with the cosmic star formation history implies that progenitors from $\sim0.1-1Z_\odot$ dominate the cosmological merging BNS population, with significant supersolar contributions. This is due to the large reservoir of star formation at relatively high metallicities ($\gtrsim 0.1\Zsun$). The precise contribution of extremely metal-poor progenitors nevertheless remains uncertain because the low-metallicity tail of cosmic star formation is poorly constrained (Section \ref{subsec:cosmic_contribution}).
    
    \item Our population models  that are best representative of the Galactic DNS sample broadly reproduce the local merger rate constraints from GWTC-5, though they typically lie at the upper end of the range $5.1-154.7\,\mathrm{Gpc}^{-3}\mathrm{yr}^{-1}$, demonstrating that massive binary evolution models capable of reproducing the Galactic DNS population need not be in tension with emerging GWTC constraints (Section \ref{subsec:galactic_dns_merger_rate}). We provide fits to the low-redshift ($z<1$) evolution of the BNS merger rate from our models (Table \ref{tab:rate_fits}), which can be tested with sGRB and kilonovae surveys and future GW observations.
    
    \item Metallicity-driven binary evolution leads to more intrinsic asymmetric mass ratio BNS mergers at high metallicities, where strong supergiant winds widens the orbit and strips the primary progenitor's envelope leading to lower mass cores (Section \ref{subsec:asymmetric_q}). This leaves more massive primary pre-SN cores among the merging BNS population (Figure \ref{fig:pre-SN_core_mass}). Meanwhile, the second-born NS is formed through ultra-stripped and electron-capture supernova, restricting its pre-SN core masses to a relatively narrow range across metallicities. Our current prescriptions connecting the pre-SN core masses to remnant NS mass do not allow us to robustly capture a significant metallicity dependence of the total BNS or individual NS mass distributions (Section \ref{subsec:mass_spectra}).
    
    \item Our observationally calibrated models predict cumulative BNS detection rates of $\sim10^4-10^5$ mergers per observer year out to $z\sim2$ in next-generation GW observatories (Section \ref{sec:detectability}). Among the proposed designs we compared, the 40\,km\,+\,20\,km configuration for Cosmic Explorer and the 2L 15\,km geometry of Einstein Telescope yields higher detection rates, with their gains growing at $z \gtrsim 1$ (Figure \ref{fig:detectability}).
\end{enumerate}

Several uncertainties should be kept in mind when interpreting these results. First, our results are sensitive to the adopted stellar wind prescriptions. In particular, the strong supergiant winds responsible for suppressing the merging \COcore channel at high metallicities remain uncertain and may be overestimated \citep{Wen+2024, Romagnolo+2026}, although we find it may be less consequential for progenitors in the NS-forming mass range compared to the more massive BH progenitors \citep[in agreement with][]{Merritt+2026}. Second, the radial expansion of stars within our grids are sensitive to the treatment of mixing efficiency \citep[e.g.,][]{Schootemeijer+2019}, thereby affecting the relative fraction of systems populating the \COcore and \Hecore channels. 
Third, the behavior of electron-capture supernovae across metallicity remains uncertain. This is particularly important because a substantial fraction of second-born NSs in our populations form through ECSN, and varying treatments of these explosions could modify both the formation efficiency and mass distribution of merging BNSs.
Finally, we do not include DNS formation through accretion-induced collapse, which may provide an additional formation pathway at high redshifts \citep{Pellouin+2025}. 

Despite these theoretical uncertainties, the overarching trends in DNS populations across cosmic history identified in this work, notably the bifurcation of the CE formation channel, the universal preference for metal-poor environments, and the shift towards more asymmetric mass ratio mergers at high metallicities, are fundamental consequences of massive binary evolution. As we transition into the era of future GW observatories and next-generation transient surveys, the anticipated number of high-redshift BNS detections will soon subject these metallicity-dependent predictions to observational tests. The models described in this study can be directly used to test whether a single model of isolated binary evolution can describe DNS formation from the Milky Way to the early Universe.

\begin{acknowledgments}
The authors acknowledge Anya Nugent, Francesco Iacovelli
and Alexandra Guerrero for helpful discussions. We also thank Sasha Levina for pointing out an issue in our \texttt{IllustrisTNG} setup. The \posydon project is supported at Northwestern University by the Gordon and Betty Moore Foundation (grant awards GBMF8477 and GBMF12341), and at the University of Geneva by the Swiss National Science Foundation (grant awards CRSII5\_213497 and 2000-1-243011) and the Foundation for the University of Geneva (FUNIGE; grant award SWISS-ET). This research was enabled by the computational resources and staff contributions provided by the Quest high-performance computing facility at Northwestern University, which is jointly supported by the Office of the Provost, the Office for Research, and Northwestern University Information Technology, hosting the Trident computing allocation (funded by the GBMF8477 and GBMF12341 grants) and the CIERA computing allocation (funded by CIERA). The computations in this paper were also performed at the University of Geneva using the Yggdrasil HPC services, and at the University of Florida using the HiPerGator supercomputing cluster provided by UFIT Research Computing (\url{http://www.rc.ufl.edu}).
M.Z. gratefully acknowledges funding from the Brinson Foundation in support of astrophysics research at the Adler Planetarium.
E.Z. is supported by a fellowship from the 'la Caixa' Foundation  (ID 100010434), code  LCF/BQ/PI26/12120016. D.R.A.-D. acknowledges funding from the Lamat Institute \citep{2025NatAs...9.1770Q} and UC Santa Cruz, supported by the Heising-Simons Foundation. 

\end{acknowledgments}

\software{This manuscript has made use of the following: 
\texttt{numpy} \citep{2020NumPy-Array},
\texttt{scipy} \citep{scipy},
\texttt{pandas} \citep{pandas},
\texttt{matplotlib} \citep{matplotlib},
\texttt{astropy} 
\citep{astropy_I, astropy_II, astropy_III},
\texttt{gwfast} \citep{gwfast2022a, gwfast2022b}.
}

\appendix

\begin{figure}
    \centering
    \includegraphics[width=0.5\linewidth]{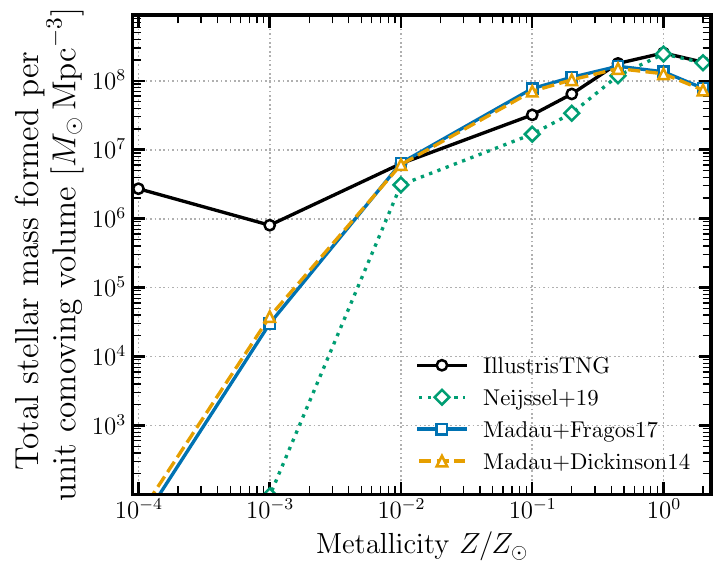}
    \caption{The total stellar mass formed per unit comoving volume integrated over cosmic history as a function of birth metallicity, calculated using Equation \ref{eq:sf_per_met}. We compare predictions from the cosmological simulation \texttt{IllustrisTNG} with three empirical prescriptions from \citet{Madau&Fragos2017}, \citet{Madau&Dickinson2014}\footnote{The \citet{Madau&Dickinson2014} star formation rate density, derived for a \citet{Salpeter1955} IMF, is multiplied by 0.66 to convert it to the \citet{Kroupa2001} IMF assumed in our populations \citep[following][]{Briel+2022}.} and \citet{Neijssel+2019} (only the first from this list and \texttt{IllustrisTNG} were adopted in this work). While all models agree that the bulk of stellar mass in the Universe is assembled at $Z > 10^{-2}\,Z_\odot$, they diverge drastically at lower metallicities. Specifically, \texttt{IllustrisTNG} predicts orders of magnitude more star formation at $Z \lesssim 10^{-2}\,Z_\odot$ compared to the sharp cutoffs inherent to the analytical models. The divergence at these low metallicity regimes constitutes a major systematic uncertainty when calculating the cosmological rates of BNS mergers originating from the most metal-poor progenitors.}
    \label{fig:total_Msun_by_met}
\end{figure}

\begin{deluxetable*}{cccccccccc}
\tabletypesize{\footnotesize}
\tablecolumns{10}
\tablewidth{1pt}
\setlength{\tabcolsep}{5pt}
\tablecaption{Parameter ranges for the supplementary \texttt{CO-HMS\_RLO} grid slices across metallicities. \label{tab:dense_CO-HMS_RLO_grid}}
\tablehead{
  \colhead{} & \multicolumn{2}{c}{Donor star} & \colhead{} & \multicolumn{2}{c}{Neutron star} & \colhead{} & \multicolumn{2}{c}{Orbit} & \colhead{} \\
  \cline{2-3} \cline{5-6} \cline{8-9}
  \colhead{Metallicity} & \colhead{$M_1$~[\Msun]} & \colhead{$\Delta\log_{10}M_1$} & \colhead{} & \colhead{$M_2$~[\Msun]} & \colhead{$\Delta M_2$} & \colhead{} & \colhead{$P_{\rm orb}\,[{\rm day}]$} & \colhead{$\Delta\log_{10}P_{\rm orb}$} & \colhead{$N$\,\tablenotemark{a}}
}
\startdata
$10^{-4}\Zsun$ & $10^{0.78}-10^{1.18}$ & 0.021 & & $1.1-2.0$ & 0.1 & & $10^{2.60}-10^{3.50}$ & 0.064 & 3000\\
$10^{-3}\Zsun$ & $10^{0.76}-10^{1.16}$ & 0.021 & & $1.1-2.0$ & 0.1 & & $10^{2.50}-10^{3.40}$ & 0.064 & 3000\\
$10^{-2}\Zsun$ & $10^{0.76}-10^{1.16}$ & 0.021 & & $1.1-2.0$ & 0.1 & & $10^{2.55}-10^{3.45}$ & 0.064 & 3000\\
$0.1\Zsun$ & $10^{0.75}-10^{1.05}$ & 0.021 & & $1.1-2.0$ & 0.1 & & $10^{2.60}-10^{3.50}$ & 0.064 & 2250\\
$0.2\Zsun$ & $10^{0.76}-10^{1.16}$ & 0.021 & & $1.1-2.0$ & 0.1 & & $10^{2.65}-10^{3.55}$ & 0.064 & 3000\\
$0.45\Zsun$ & $10^{0.76}-10^{1.16}$ & 0.021 & & $1.1-2.0$ & 0.1 & & $10^{2.65}-10^{3.55}$ & 0.064 & 3000\\
$\Zsun$ & $10^{0.76}-10^{1.20}$ & 0.020 & & $1.1-2.0$ & 0.1 & & $10^{2.70}-10^{3.60}$ & 0.064 & 3300\\
$2\Zsun$ & $10^{0.78}-10^{1.18}$ & 0.021 & & $1.1-2.0$ & 0.1 & & $10^{2.75}-10^{3.65}$ & 0.064 & 3000\\
\enddata
\tablenotetext{a}{Number of models in this metallicity grid.}
\end{deluxetable*}

\bibliography{sample701}{}
\bibliographystyle{aasjournalv7}

\end{document}